\documentclass[12pt]{article}
\pdfoutput=1

\usepackage{epsf}
\usepackage{epsfig,graphics}
\usepackage {graphicx}
\usepackage {epsfig}
\usepackage {subfigure}
\usepackage {tabularx} 
\usepackage{rotate}	
\usepackage{slashed}

\usepackage{amsmath}
\usepackage{amsfonts}
\usepackage{amssymb}
\usepackage{graphicx}
\usepackage{cite}
\usepackage{float}

\usepackage{fancyhdr}
\usepackage{hyperref}

\newcommand{\bmat}{\left(\begin{array}}
\newcommand{\emat}{\end{array}\right)}

\def\yzero{\smash{\hbox{$y\kern-4pt\raise1pt\hbox{${}^\circ$}$}}}

\def\beq{\begin{equation}}
\def\eeq{\end{equation}}
\def\beqa{\begin{eqnarray}}
\def\eeqa{\end{eqnarray}}

\def\-{\hphantom{-}}

\def\s2{\frac{1}{\sqrt2}}

\def\beq{\begin{equation}}
\def\eeq{\end{equation}}
\def\beqa{\begin{eqnarray}}
\def\eeqa{\end{eqnarray}}

\def\IF{\relax{\rm I\kern-.18em F}}
\def\II{\relax{\rm I\kern-.18em I}}
\def\IP{\relax{\rm I\kern-.18em P}}
\def\IC{\relax\hbox{\kern.25em$\inbar\kern-.3em{\rm C}$}}
\def\IR{\relax{\rm I\kern-.18em R}}

\def\Dsl{\,\raise.15ex\hbox{/}\mkern-13.5mu D} 
\def\IZ{Z\kern-.4em  Z}

\catcode`\@=11   
\newdimen\@rotdimen
\newbox\@rotbox  

\def\@vspec#1{\special{ps:#1}}
\def\@rotstart#1{\@vspec{gsave currentpoint currentpoint translate
   #1 neg exch neg exch translate}}
\def\@rotfinish{\@vspec{currentpoint grestore moveto}}
\def\@rotr#1{\@rotdimen=\ht#1\advance\@rotdimen by\dp#1%
   \hbox to\@rotdimen{\hskip\ht#1\vbox to\wd#1{\@rotstart{90 rotate}%
   \box#1\vss}\hss}\@rotfinish}
\def\@rotl#1{\@rotdimen=\ht#1\advance\@rotdimen by\dp#1%
   \hbox to\@rotdimen{\vbox to\wd#1{\vskip\wd#1\@rotstart{270 rotate}%
   \box#1\vss}\hss}\@rotfinish}%
\def\@rotu#1{\@rotdimen=\ht#1\advance\@rotdimen by\dp#1%
   \hbox to\wd#1{\hskip\wd#1\vbox to\@rotdimen{\vskip\@rotdimen
   \@rotstart{-1 dup scale}\box#1\vss}\hss}\@rotfinish}%
\def\@rotf#1{\hbox to\wd#1{\hskip\wd#1\@rotstart{-1 1 scale}%
   \box#1\hss}\@rotfinish}%
\def\rotate{\@ifnextchar[{\@rotate}{\@rotate[l]}}
\def\@rotate[#1]#2{\setbox\@rotbox=\hbox{#2}\@nameuse{@rot#1}\@rotbox}

\catcode`\@=12

\topmargin
-1.5cm
\textwidth
15.5cm
\textheight
23.5cm
\oddsidemargin
0.7cm
\evensidemargin
0.7cm

\begin{document}

\makeatletter
\@addtoreset{equation}{section}
\makeatother
\renewcommand{\theequation}{\thesection.\arabic{equation}}
\pagestyle{empty}
\vspace{-0.2cm}
\rightline{ IFT-UAM/CSIC-18-032}
\vspace{1.0cm}
\begin{center}


\LARGE{AdS-phobia, the WGC, the Standard Model \\
and Supersymmetry\\
[3mm]}
  \large{Eduardo Gonzalo, Alvaro Herr\'aez and Luis E. Ib\'a\~nez \\[6mm]}
\small{
  Departamento de F\'{\i}sica Te\'orica
and Instituto de F\'{\i}sica Te\'orica UAM/CSIC,\\[-0.3em]
Universidad Aut\'onoma de Madrid,
Cantoblanco, 28049 Madrid, Spain 
\\[4mm]}
\small{\bf Abstract} \\[3mm]
\end{center}
\begin{center}
\begin{minipage}[h]{15.22cm}
It has been recently argued that an embedding of the SM into a consistent theory of quantum gravity
may imply important constraints on the mass of the lightest neutrino and the cosmological constant $\Lambda_{4}$.
The constraints come from imposing the absence of any non-SUSY AdS stable vacua obtained from any  consistent compactification
of the SM to 3 or 2 dimensions. This condition comes as a corollary of a recent extension of the
{\it Weak Gravity Conjecture} (WGC)  by Ooguri and Vafa. In this paper we study   $T^2/Z_N$ compactifications of the SM
to two dimensions  in which
SM Wilson lines are projected out, leading to a 
 considerable simplification. We analyze in detail a $T^ 2/Z_4$ compactification of the SM in which 
 both  complex structure and Wilson line scalars  are fixed and the potential is only a function of the area of the torus $a^2$.
 We find that the SM is not robust against the appearance of AdS vacua in 2D and hence would be 
 by itself  inconsistent with quantum gravity. 
 On the contrary, if the SM is embedded at some scale $M_{SS}$  into a SUSY version like the MSSM, 
 the AdS vacua present in the non-SUSY case disappear or become unstable.  This means that WGC arguments favor a
 SUSY version of the SM, independently of the usual hierarchy problem arguments.
In a
 $T^2/Z_4$ compactification in which the orbifold action is embedded into  the $B-L$ symmetry  the bounds on neutrino masses 
 and the cosmological constant are recovered.  This suggests that the MSSM should be extended with a  $U(1)_{B-L}$  gauge group.
  In other families of vacua the spectrum of SUSY particles is further  constrained in order to avoid
 the appearance of new AdS vacua or instabilities.
  We discuss a 
possible understanding  of the {\it little hierarchy problem} in this context.

\end{minipage}
\end{center}
\newpage
\setcounter{page}{1}
\pagestyle{plain}
\renewcommand{\thefootnote}{\arabic{footnote}}
\setcounter{footnote}{0}



\tableofcontents
\newpage

\section{Introduction}

In this paper we study possible constraints on Standard Model (SM) physics derived from its 
eventual embedding into a consistent theory of quantum gravity.  As first emphasised by Vafa  \cite{swampland},
many simple field theories do not admit such an embedding and populate what he called the 
{\it swampland} of theories. Since then there has been an important effort in trying to delimitate what is 
the space of theories which belong to the swampland,
see \cite{WGC,WGC1,WGC2,aunmas} for an incomplete list of references and \cite{vafafederico} for a recent review.
The most studied constraints to delimit the swampland are those given by the Weak Gravity Conjecture (WGC) 
\cite{WGC}, which loosely speaking implies that the gravitational interaction must be weaker than gauge interactions in
any consistent theory of quantum gravity. In its simplest incarnation of a $U(1)$ theory coupled to gravity it implies that 
there must be at least one charged particle with mass $m$  and charge $Q$ such that $m \leq Q$ in Planck units, and is
motivated by blackhole physics. There are different
versions of the conjecture  and  also extensions to multiple $U(1)$'s and to the antisymmetric gauge tensors of string theory
and supergravity, see \cite{WGC,WGC1,WGC2,aunmas,vafafederico}.  In the latter case it is the tension of the branes of string theory 
which are bounded in terms of the charge of the antisymmetric RR or NS tensors, so that $T\leq Q$.
Recently Ooguri and Vafa  (OV) put forward a {\it sharpened} version of the WGC \cite{OV} (see also \cite{moreOV}) stating that in non-SUSY theories
the inequality is strict, $T< Q$. When applied to string flux vacua  the OV conjecture implies that 
all consistent AdS vacua must be either SUSY or unstable.  This instability implies that there is no CFT dual  and, in this sense,
there is no consistent quantum gravity theory with a stable non-SUSY AdS vacuum. Indeed there is so far no counterexample to this 
conjecture in string theory\footnote{There are examples of non-SUSY theories with AdS/CFT duals but, unlike the SM coupled to 
gravity, there are infinite towers of higher spin particles and do not correspond to ordinary Einstein gravity with a finite number of fields,
see e.g.\cite{giombi} and references therein).}.  The authors of \cite{OV} go also beyond the
string theory setting and conjecture that non-SUSY AdS/CFT duality is in the swampland in general, and not only for flux vacua.

We will call {\it AdS-phobia} for short  the condition of the absence of any AdS stable non-SUSY vacuum in a given theory, as suggested by the
more general OV condition. AdS-phobia, if correct, would be a very powerful constraint   on physics models. For example it would drastically reduce the 
possibilities in the string landscape: only Minkowski and de Sitter  vacua would be consistent with quantum gravity and only those 
(in addition to SUSY ones) would 
in principle count in addressing the enormous multiplicity of vacua in string theory. 

It has been recently pointed out that AdS-phobia, if applied to the SM,  would have important implications on SM physics \cite{OV,IMV1,IMV2}.
The point is that, as described in \cite{ArkaniHamed:2007gg}, compactifying the SM to  3 or 2  dimensions one may obtain AdS  vacua,
depending on the values of the cosmological constant $\Lambda_{4}$ and the neutrino masses. On the other hand the assumption of
background independence implies that if a theory is consistent in 4D, all of its compactifications should be consistent also, since we have only changed the
background geometry. Thus the existence of these lower dimensional AdS vacua  (if stable) would imply the inconsistency of the 
SM itself\cite{OV,IMV1}. A detailed analysis of these constraints on neutrino masses and the cosmological constant was presented in
\cite{IMV1}.  Four interesting conclusions were obtained by imposing the absence of AdS minima,
\begin{itemize}
	\item  Majorana neutrinos (two degrees of freedom per neutrino) are not possible \cite{OV}.
	\item The lightest neutrino mass is bounded as $m_{\nu_1}\leq 4\times 10^{ -3}$ eV (normal hierarchy) or $m_{\nu_3}\leq 1\times 10^{ -3}$ eV (inverted hierarchy)\cite{IMV1}.
	\item The cosmological constant $\Lambda_4$ is bounded from below by the mass of the lightest (Dirac) neutrino, $\Lambda_4\gtrsim m_{\nu_i}^ 4$\cite{IMV1}.
	\item The upper bound on the lightest neutrino mass implies an upper bound on the Electro-Weak  scale
	(for fixed  neutrino Yukawa couplings)\cite{IMV1,IMV2}.
\end{itemize}
The last  point is obvious from the fact that the neutrino mass depends on the Higgs vev. Any increase in the EW scale above a scale $\simeq 1$ TeV would make the lightest neutrino mass to be too large and violate the bound to avoid AdS. This is remarkable because it would imply that values of the EW scale much above the observed scale would
belong to the swampland. Thus, in this sense,  the EW fine-tuning problem would be a mirage, since the EW scale is fixed in order to avoid AdS vacua,
and its scale is secretly  dictated by $\Lambda_{4}$ and the neutrino Yukawa couplings.
The smallness of the EW scale compared to the Planck scale would be related to the even greater hierarchy of the cosmological constant versus the Planck scale \cite{IMV1,IMV2}.

In \cite{IMV1} the neutrino  AdS vacua  in 3D and 2D were analyzed in detail. These minima appear from an interplay between the 
(positive) dimensional reduction of the cosmological constant term,  the (negative) Casimir contribution of the photon and graviton and the (positive) Casimir contribution of the lightest fermions of the SM, the neutrinos. If the lightest neutrino is
sufficiently light, its contribution can destroy the AdS vacuum and make the SM AdS-safe. This is the origin of the bound on the
lightest neutrino. The Casimir contribution of the other SM particles is exponentially suppressed with their mass.  However as one explores the radion potential (e.g. in 3D in a circle) at shorter radius R, the thresholds of the other SM particles are reached and their contribution needs to be taken into account. So the question arises whether there are further local minima 
or runaway directions 
at shorter radii. In particular, as one goes up in energies (smaller $R$) additional scalar moduli appear associated first with the Wilson line of the 
photon and then, above the hadron  threshold, two Wilson lines in the Cartan subalgebra of QCD. At even shorter distances the Wilson line of the $Z^0$ 
appears. So the full moduli space in 3D involves one radion and four Wilson lines, with a scalar potential depending on all of them.  An analysis of the existence of
these additional Wilson line dependent minima was recently carried out in \cite{HS}, concluding that indeed runaway field directions at smaller R exist for
non-trivial photon Wilson line. The $R\simeq 1/m_\nu$ AdS vacua are bona-fide local minima but if they could tunnel  into these field directions 
there would be no constraints on neutrino masses nor the EW hierarchy.

In this paper we reexamine and extend the study of the AdS minima of the SM both to smaller compact radii
 and geometries other than the circle and the torus.  It is useful to classify the vacua we find in three categories:
 \begin{itemize}
\item {\it Type D (Dangerous)}. Vacua which contain a stable AdS minimum which cannot be avoided by constraining free parameters of the theory. 
\item {\it Type S (Safe)}.  Vacua which contain  no stable AdS minima.
\item {\it Type P (Predictive)}.  Vacua which contain no stable AdS minima for a certain range of the free parameters of the theory.
 \end{itemize}
 Our attitude concerning instability will be conservative in the sense that we will consider unstable any potential which includes any runaway direction in the fields,
 even though we are not certain how the tunneling from local minima to those runaway directions could take place in detail.
 We find minima in the three categories.
After examining the case of the compactification on a circle, we discuss SM compactifications in which Wilson line moduli are fixed.
After a brief discussion of the compactification on the segment
$S^1/Z_2$,  we discuss the compactification on toroidal orbifolds $T^2/Z_N$.
We  specialize to the case of the $T^2/Z_4$ orbifold, but most implications apply to $Z_{N}$, $N=2-6$ in general.
 The reason to study this particular  background is twofold. On the one hand, unlike the case of the parity reflection in the segment, a $Z_4$ rotation
 is a symmetry of the uncompactified 4D SM. 
 Secondly,  in such a background all the Wilson lines as well as the complex structure are
fixed, so that the scalar potential depends  only on the torus area.  This simplifies enormously the study of the scalar potential at
shorter distances  and runaway directions induced by Wilson line moduli disappear. It turns out that,
in order to obtain predictive $T^2/Z_4$ vacua one needs to embed the discrete rotation into internal symmetries of the SM. 
We find that different embeddings into the SM  gauge group lead necessarily to stable AdS vacua, rendering the SM inconsistent with quantum gravity.
This is true irrespective of the value of neutrino masses or any other free particle physics parameter.  
Thus some of the SM vacua are of {\it Type D} and make the SM as such not 
viable.

The underlying reason why the SM has these AdS vacua is that it is {\it too fermionic}, it has many more fermions than bosons. The dangerous
AdS vacua could become unstable {\it Type S} if there were enough bosons to compensate in the Casimir potential.
A situation in which this naturally happens
is a SUSY embedding of the SM like the MSSM. Thus, unlike the SM, the MSSM seems robust against the generation of stable AdS vacua to the extent that
we have not found vacua of  {\it Type D}.
Furthermore we find $Z_4$ embeddings into a discrete subgroup of the $U(1)_{B-L}$ gauge symmetry 
leading to AdS vacua of {\it Type P}  which may be evaded for appropriate neutrino masses, like the
original 3D vacua in the circle, and hence are predictive. This suggests that the MSSM should be extended at some scale with an
extra $U(1)_{B-L}$ gauge symmetry.
 Another class of embeddings into discrete R-symmetries,  in addition to neutrinos,  also constraint the masses
of the SUSY particles in a way essentially dictated by the sign of  the  supertrace $\delta_{SS}=\sum_i(-1)^{(n_i)}m_i^2$.
More generally,  in SUSY models avoiding AdS vacua  from charge-colour breaking AdS minima inherited from the 4D potential impose additional constraints.
Thus an AdS-safe MSSM should pass both these constraints, leading potentially  to interesting conditions for SUSY model building.

As discussed in \cite{IMV1,IMV2}, avoiding the AdS neutrino vacua imply an upper bound on the EW scale close to its experimental value.
On the other hand absence of lower 4D charge-colour breaking minima requires typically a relatively massive SUSY spectrum. This could 
provide for an explanation for the {\it little hierarchy problem} of the MSSM:  absence of AdS neutrino vacua requires a EW scale close to its experimental value, 
while absence of charge-colour breaking AdS minima pulls up the value of SUSY masses.  

It is important to remark that the  absence of consistent AdS non-SUSY vacua is at the moment only a conjecture, although in
agreement with the string examples examined up to now.  Furthermore, the results rely on the assumption that the obtained AdS SM vacua
are stable.  Although indeed the obtained  minima appear stable, if the theory e.g. is embedded into a string landscape, one cannot rule
out transitions into other lower energy vacua which might exist in the landscape. This issue is discussed  in \cite{IMV1}.
It  is difficult to quantify, given our ignorance of the structure of the landscape.
In the present paper we assume that those transitions are absent and the AdS minima are stable, 
and explore what the physical consequences would be.

The structure of this paper is as follows. In the next section we review the case of the compactification of the SM  on a circle and the 
origin of the corresponding bounds on neutrino masses and the cosmological constant, exploring also values of the radii up to the EW scale.
 We then consider the structure of
the potentials at small $R$  with fixed Wilson lines, as would appear in a 
compactification on the segment $S^1/Z_2 $.  In section \ref{sec:SMtorusorbifold} we discuss the compactification of the SM on  $T^2/Z_4$
and show how the SM has vacua of {\it Type D} and hence would be incompatible with quantum gravity. In section \ref{sec:torusMSSMvacua} 
we show how such vacua are not present in the SUSY case and how one recovers the neutrino mass bounds if
$U(1)_{B-L}$ (or a discrete subgroup) is gauged at some level.
New vacua may also lead to constraints on the SUSY spectrum.
 We leave section \ref{sec:conclusions}   for our conclusions and outlook. In Appendix \ref{app:1loopcircle} we discuss the computation of the 
Casimir potential in the SM on the circle, including also the EW degrees of freedom. In Appendix \ref{app:1looptorusorbifold} we work out some  results for the
compactification on the $T^2/Z_4$ orbifold. Appendix \ref{app:minkowski} contains a discussion on the consistency of assuming a flat non-compact background 
in the computation of Casimir energies. Finally, Appendix \ref{app:MSSMspectrum}
 contains a table with the MSSM spectrum used to draw
some of the plots.

\section{The SM in 3D}

In this section we first study the compactification of the SM coupled to Einstein gravity with a cosmological constant 
down to 3D on the circle. We begin by exploring the regions with the compact radius larger than the electron wave length 
and then regions with smaller radius, going up to the EW scale.  The resulting theory has one radius, one Higgs and four Wilson line scalar variables.
We reproduce the local  minima associated to the neutrino region and study the potential at shorter radii. The potential features 
runaway  directions in the presence of Wilson line moduli that may lead to decay of the neutrino vacua through tunneling. 
We also discuss the structure of the radius effective potential in the case with frozen Wilson lines. This would happen 
in particular in theories compactified in the segment $S^ 1/Z_2$ and serves as an appetizer to the case of compactifications
on $T^2/Z_N$ which are discussed in the next section.  In this case it is easier the  search for minima for the radius potential,
since then all Wilson lines are fixed.

\subsection{The SM on the circle I ($R\gtrsim 1/m_{e}$)}

We first  briefly summarize and extend some of the results in \cite{ArkaniHamed:2007gg, IMV1} so that the reader better understands the further results 
presented in this paper.
In \cite{ArkaniHamed:2007gg} the compactification of the SM Lagrangian to lower dimensions was considered.  One of the purposes
of that work was to show how the notion of landscape of vacua should not be associated exclusively to string theory,  but even the SM
itself has a wealth of vacuum solutions when compactified to $3D$ or $2D$. They concentrated in the deep infrared sector of the SM, below the 
electron threshold, in which the only relevant particles are the photon, the graviton and the neutrinos. They found that, depending on the value
of the neutrino masses, the SM possesses local minima both on 3D and 2D, with very large compactification radii of order $R\simeq 1/m_\nu$.
The appearance of these minima goes as follows.  The action of the pure gravity action reduced to 3D in a circle is given by 
\begin{eqnarray}
S_{GR}= \int d^3x \sqrt{-g_3} (2\pi)\left[\frac{1}{2}M_p^2\mathcal{R}_{(3)}-\frac{1}{4}{R^4}W_{\mu\nu}W^{\mu\nu}-M_p^2\left(\frac{\partial R}{R}\right)^2 - \frac{\Lambda_4}{R^2}\right].
\label{eq:action}
\end{eqnarray}
Here $M_p$ is the 4D reduced Planck mass, $M_p=(8\pi G_N)^{-1/2}$ and $\Lambda_4$ is the 4D cosmological constant.
The action of the graviphoton of field strength $W_{\mu\nu}$ is also shown. There is a runaway potential inherited from the 4D cosmological term. Nevertheless, the cosmological constant is so small that the quantum contribution of the lightest SM modes to the effective potential becomes relevant. Their contribution at one-loop level can be identified with the Casimir energy, which for a non-interacting massless field with periodic boundary conditions is found to be:
\begin{equation}
V_{p}(R) \ =\ \mp \frac{n_p}{720 \pi}\frac{1}{R^6},
\end{equation}
where $n_p$ counts the number of degrees of freedom of the particle.
For bosons the negative sign applies, whereas for fermions the sign is positive.
The only massless states in the SM are the photon and the graviton, so in the absence of any other light particle 
it is clear that the potential becomes unstable and negative for small $R$.  However, if there are fermions with a mass of order 
$m\gtrsim \Lambda_4^{1/4}$, minima can develop. The Casimir potential for a non-interacting massive particle of spin $s_{p}$ and mass $m_{p}$ has the form \cite{ArkaniHamed:2007gg}
\begin{equation}
V_{p}(R)=(-1)^{2s_{p}+1}n_{p}\frac{m_{p}^{2}}{8\pi^{4}R^{4}}\sum_{n=1}^{\infty}\frac{K_{2}(2\pi nR m_{p})}{n^{2}}.
\end{equation}
Here $K_{2}(x)$ is a modified Bessel function of the second kind.  This gives us the leading quantum contribution to the radion potential.
As long as couplings remain perturbative, higher loop corrections will be small and unimportant. This excludes the region close to the
QCD scale in which non-perturbative techniques would be appropriate. We will thus concentrate in the potential above or below the QCD scale. 
In the latter case we will count pions and kaons as elementary.  

The Bessel function decays 
exponentially for large argument. For this reason, to work out the structure of the potential around the scale $\Lambda_4^{1/4}$  within the SM,  only the neutrinos will be relevant, since heavier particles like the electron contribute in a negligible way to the local potential.  Neutrino oscillation experiments tell us what are the mass
differences between  neutrino masses, but not their Dirac or Majorana character. We do not know whether the hierarchy of the masses is normal (NH) or
inverted (IH).  Experimental constraints tell us  for the neutrino mass differences that \cite{Olive:2016xmw}
\begin{eqnarray}
\Delta m_{21}^2 = (7.53\pm 0.18)\times 10^{-5}\,\, {\rm eV}^2,\\
\Delta m_{32}^2 = (2.44\, \pm 0.06)\times 10^{-3}\,\, {\rm eV}^2 \,\,{\rm (NH)},\\
\Delta m_{32}^2 =(2.51\pm 0.06)\times 10^{-3}\,\, {\rm eV}^2 \,\,{\rm (IH)}.
\end{eqnarray}
We do not know what the mass of the lightest neutrino is, and it is not experimentally excluded that it could be massless.
We use the above neutrino constraints in computing the radion potential and plotting the figures.
We always plot the potential divided by the contribution of a single massless degree of freedom, so that we can interpret it as the number of effective degrees of freedom.
Thus, the potential in our plots appears multiplied by $R^{6}$ times a certain constant. Care
must be taken when extracting physical conclusions  from these plots,
since minima of $R^{6}V$ need not be minima of $V$. However,
an AdS minima of $R^{6}V$ which eventually turns positive (in
all directions), always corresponds to an AdS minima of $V$.

\begin{figure}[t]
     \begin{center}
     	\subfigure[]{	
        \includegraphics[scale=0.36]{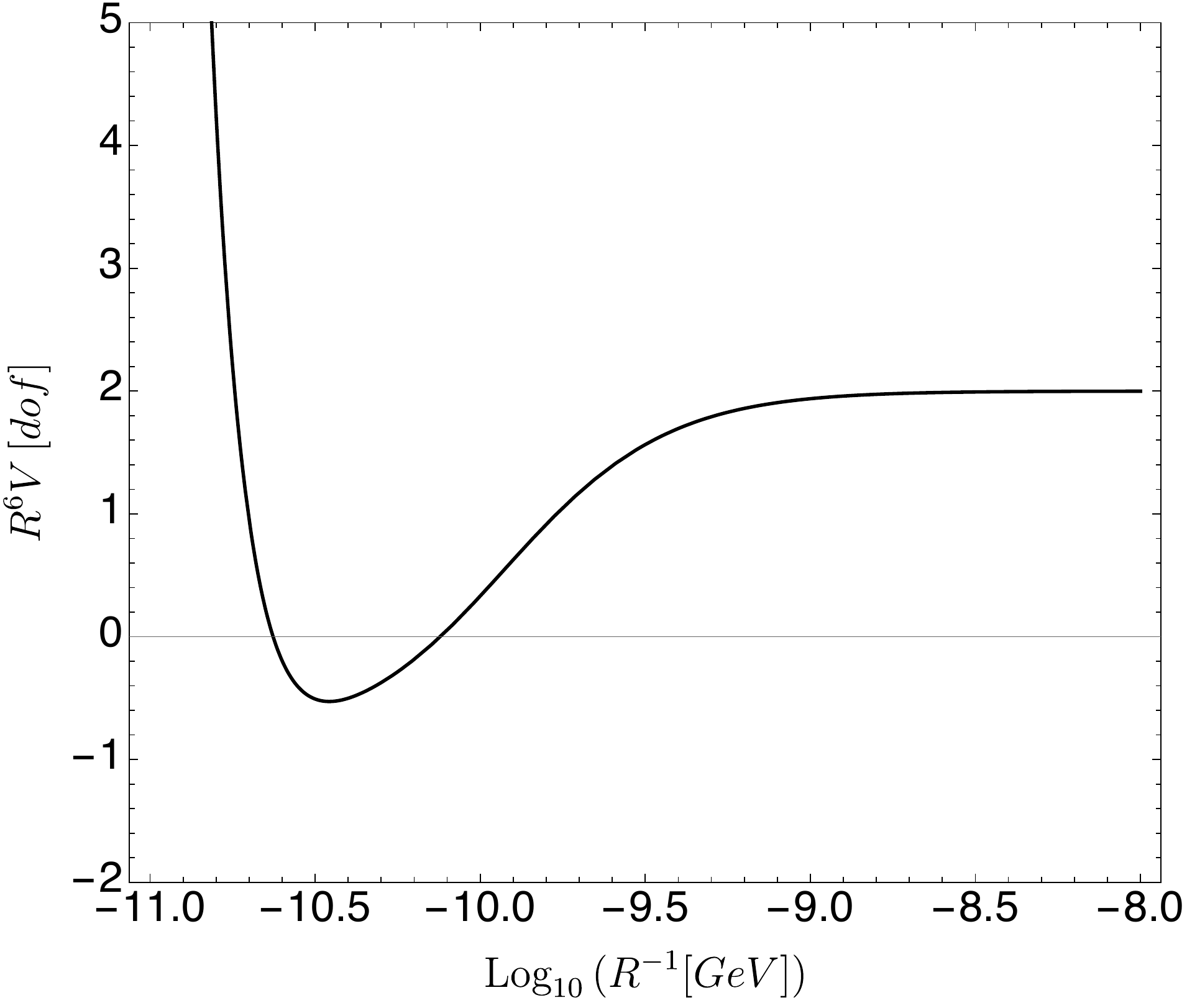}
        \label{majofigleft}
        }
        \subfigure[]{
        \includegraphics[scale=0.355]{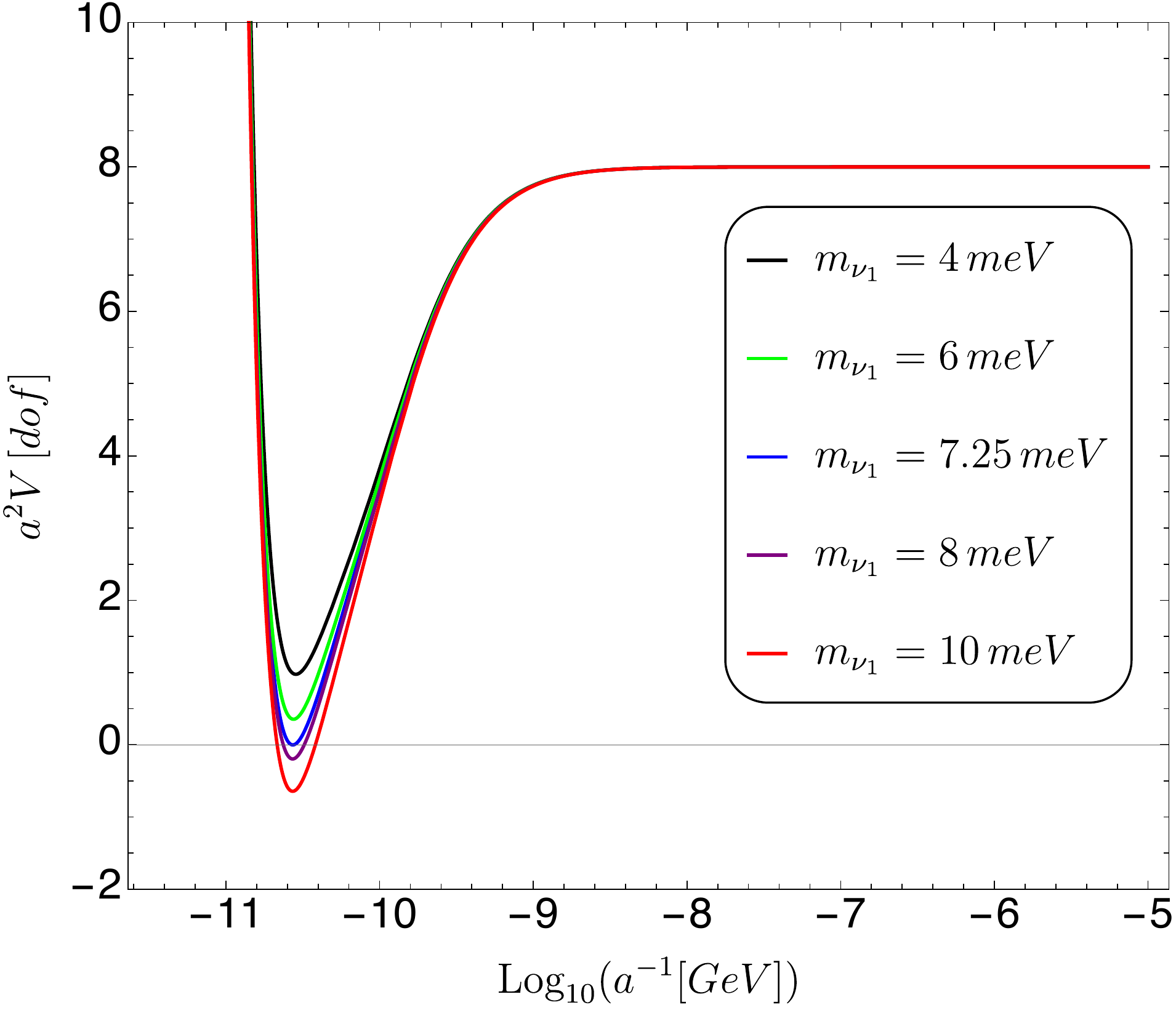}
        \label{majofigright}
        }
      \caption{\footnotesize  \textbf{(a)} Radion effective potential for Majorana neutrinos. The lightest neutrino is considered massless and a minimum develops.  For higher energies, the neutrinos behave as massless particles. We have 4 bosonic and 6 fermionic degrees of freedom so, in the massless limit, $R^{6}V$ (in units of degrees of freedom) tends to 2. \textbf{(b)} Radion effective potential for Dirac neutrinos. The different lines correspond to several values for the lightest neutrino mass $m_{\nu_1}$. We have 4 bosonic and 12 fermionic degrees of freedom so, in the massless limit, $R^{6}V$  tends to 8. It is important to remember that positive valued  minima in this plot do not necessarily correspond to dS minima of $V$. However, negative minima do correspond to AdS minima of $V$.}
        \label{majofig}
     \end{center}
\end{figure}

If neutrinos are Majorana and are given periodic boundary conditions, we have six fermionic degrees of freedom which at small $R$ dominate over the 
photon/graviton  four degrees of freedom, so that the potential grows. However when going to values $R\simeq 1/m_{\nu_1}$ an AdS minimum is always created,
irrespective of the neutrino masses, 
since only two neutrino degrees of freedom can become sufficiently light and those cannot overwhelm the bosonic contribution at large R, see Fig. \ref{majofigleft}.
On the other hand if neutrinos are Dirac, up to 4 neutrino degrees of freedom may be sufficiently light to compensate for the four bosonic ones.
In this case minima forms or not depending on the value of the lightest neutrino mass, see Fig. \ref{majofigright}.   

If we apply the principle of no AdS non-SUSY vacua we reach the conclusion that  neutrino Majorana masses are not possible and we obtain bounds
for the lightest neutrino Dirac mass  $m_{\nu_1}>7.7\times 10^{-3}$ eV for (NI) and $m_{\nu_3}>2.56\times 10^{-3}$ eV for (IH), see \cite{IMV1} for details
\footnote{Note that what counts is that the number of degrees of freedom of the lightest neutrino should be  4. Thus it may also be {\it pseudo-Dirac}, 
i.e. it may have both Dirac and Majorana mass contributions of the same order of magnitude. In this case $\nu$-less double $\beta$-decay could be posible.}.

As we mentioned before, another interesting implication is that within this scheme there is a lower bound on the value of the  4D cosmological constant \cite{IMV1}, qualitatively
$\Lambda_4 \gtrsim m_{\nu_1}^4$.  This is interesting because  it is the first argument implying a non-vanishing $\Lambda_4$ only on the basis of 
particle physics, with no cosmological input.

A further implication concerns the fine-tuning issue of the  Electro-Weak (EW) scale in the SM. Indeed, the mentioned upper bounds on neutrino masses
imply an upper bound on the Higgs vev, for fixed Yukawa coupling. Thus SM versions with a Higgs sector leading to Higgs vevs above 1 TeV would be in the swampland,
since larger vevs implies larger neutrino masses and hence AdS vacua would form in 3D or  2D  \cite{IMV1}. In particular for Dirac neutrino masses one has 
a bound
\beq 
\left|\left\langle H\right\rangle \right| \ \lesssim \ 1.6(0.4)\frac {\Lambda_4^{1/4}}{Y_{\nu_1}} \ ,
\label{jerarquia}
\eeq
for NH(IH) neutrinos.  Here $Y_{\nu_1}$ is the lightest neutrino Yukawa coupling and $\Lambda_4$ the 4D cosmological constant.
With values of $\Lambda_4$ as observed in cosmology, larger values for the EW scale would necessarily lead, for fixed Yukawa, to undesired AdS minima. 
From the low-energy Wilsonian point of view the smallness of the EW scale looks like an enormous fine-tuning, but this would be a mirage since  field theory
parameters leading to larger EW scale would not count as possible consistent theories.  The smallness of the EW scale compared to the Planck scale is here
tied up to the smallness of the cosmological constant. Note that here we are taking the value of  $\Lambda_4$ fixed to its observed value while varying $H$. 
This may be motivated by the fact that anthropic arguments from galaxy formation require  $\Lambda_4^{1/4}\simeq 2.25\times 10^{-3}eV$ within a factor of order 4 \cite{anthropic}.
So we we vary $H$ while keeping the cosmological constant around its measured present value.

Having Dirac neutrinos rather than Majorana departs from the most common scenario to understand the smallness of neutrino masses, the seesaw
mechanism. In the present approach the lightest neutrino is very light to avoid that an AdS vacuum forms. Its mass is bounded from above by 
$m_{\nu_1}\lesssim \Lambda_4^{1/4}$, giving an explanation for the apparent numerical coincidence of neutrino and cosmological constant scales.
Strictly speaking, it is only necessary for one of the three neutrinos to be that light. However, if neutrino Yukawa couplings of the different generations
are correlated, as they are for the charged leptons and quarks, this condition will drag all of the neutrinos to be very light, as observed
\footnote{Note in this respect that obtaining Majorana neutrino masses in string compactifications has proven to be a notoriously difficult task both in
heterotic and Type II orientifold constructions.  In the latter, Majorana neutrino masses may in principle appear from  string instanton effects
\cite{instantneutrinos, Blumenhagen:2006xt}. However, a dedicated search \cite{isu}  for such instanton effects in a large class of  MSSM-like Gepner orientifolds was unsuccesful.
On the other hand  $\nu_R$ Majorana masses may appear from non-renormalizable couplings   of right-handed neutrinos to scalars breaking the 
$B-L$ symmetry in extended gauge symmetry models.}. Still, one would  like to have in addition an explanation to why the neutrino Yukawa couplings are so small.
As emphasized in \cite{IMV1}, neutrino Majorana masses are still allowed if we go beyond the SM and assume there is an additional singlet Weyl fermion  $\chi$ with mass
$m_\chi\lesssim 10^{-3}$ eV.  Then the AdS vacua is again avoided if the lightest  (Majorana) neutrino is light enough. In this case the bound on the EW symmetry scale 
has the form \cite{IMV1,IMV2}
\beq
\frac {(Y_{\nu_1}\left\langle H\right\rangle )^2}{M} \ \lesssim\ 2\times \Lambda_4^{1/4} \ \longrightarrow \  \\
\left\langle H\right\rangle  \ \lesssim  \ \frac {\sqrt{2}}{Y_{\nu_1}} \sqrt{M \Lambda_4^{1/4}}  \ .
\label{hierarchy}
\eeq 
where $M$ is the scale of lepton number violation in the see-saw mechanism. 
In this case the EW scale  is bounded from above by
the geometric mean of the cosmological constant scale and the lepton number violation scale $M$. Thus, e.g. for $Y_{\nu_1}\simeq 10^{-3}$ and
$M\simeq 10^{10}-10^{14}$ GeV, one gets $\left\langle H\right\rangle \lesssim  10^2-10^4$ GeV.  This possibility is interesting because, since it uses the see-saw 
mechanism, no hierarchically small Yukawas are needed.  In the rest of this paper it will not be relevant whether   neutrinos are 
Dirac or Majorana +$\chi$, but we think it is interesting to keep in mind the different possibilities.

All these constraints obtained imposing the absence of AdS local minima  assume that these vacua are absolutely stable.
A possible source of instabilities may in principle arise if there are lower minima or runaway directions at smaller radion values (higher thresholds)
contributing to the Casimir potential \cite{HS}.  This is an important motivation to go to  the orbifold case, as we will discuss in section \ref{sec:SMtorusorbifold}.

\subsection{The SM on the circle II ($1/m_e\gtrsim R\gtrsim1/M_{EW}$)}

In this section we study the one-loop effective potential of the full
Standard Model with massive Dirac neutrinos and minimal coupling to gravity
compactified in a circle. Our calculations were carried out using
the Background Field Method, in the spirit of \cite{Appelquist}.
 Some details on the computations can be found in  Appendix \ref{app:1loopcircle}.
We parameterize the metric as in the usual Kaluza-Klein ansatz:
\begin{equation}
g_{\mu\nu}=\frac{1}{R^{2}}\left[\begin{array}{cc}
g_{ij}+R^{4}B_{i}B_{j} & R^{4}B_{i}\\
R^{4}B_{j} & R^{4}
\end{array}\right].
\end{equation}
We will set the graviphoton $B_{i}=0$ from the start, since its vev is zero and it does not contribute to the effective potential. The scalar fields in the 3D Effective Action are the radion $R$,
the Higgs field $H$ (whose vev is $\text{v}\simeq246$ GeV) and the
Wilson lines of all the gauge bosons in the Cartan Subalgebra of $SU(3)_{C}\times SU(2)_{L}\times U(1)_{Y}$:
two gluon Wilson lines ($G_{1}$,$G_{2}$), the photon $A$ and the
$Z$ boson. To perform the compactification we need to expand our
fields $\phi$ in a basis that respects the boundary condition:
\begin{equation}
\phi(x_{i},y+2\pi)=e^{i2\pi\left(\frac{1-z}{2}\right)}\phi(x_{i},y)\text{},
\end{equation}
since $y\in\left(0,2\pi  \right)$. Fermions can have either periodic
$z=1$ or antiperiodic $z=0$ boundary conditions, while bosons are
only allowed to have periodic ones. We use Fourier harmonics:
\begin{equation}
\phi(x_{i},y)=\sum_{n=-\infty}^{\infty}\phi_{n}(x_{i})e^{i2\pi\left(n+\frac{1-z}{2}\right)y}.
\end{equation}
The relevant classical contribution to the effective potential in three dimensions
consists of a term coming from the 4D cosmological constant, the Higgs potential and a mass term for the $Z^0$ boson Wilson line  ($Z$).
 We will neglect the effects of a non-vanishing 3D curvature in the Casimir computation.
We argue in Appendix  \ref{app:minkowski} that those effects may be safely neglected as long as the radii are not exceedingly small.
The one-loop potential can be consistently identified with
the Casimir energy $V_{p}$ of the different particles. One then has
\begin{align}
V[R,H,Z,A,G_{1},G_{2}] & =\frac{1}{R^{2}}\biggl[2\pi\  \Lambda_{4}+\frac{1}{2}m_{H}^{2}(H-\text{v})^{2}+\lambda\text{v}(H-\text{v})^{3}+\frac{\lambda}{4}(H-\text{v})^{4}\nonumber \\
& \quad+\frac{1}{2}M_{Z}^{2}\frac{H^{2}}{\text{v}^{2}}Z^{2}\biggl]+\sum_{p}V_{p}.\label{eq:globalformula}
\end{align}
As explained in  Appendix \ref{app:1loopcircle}, the contribution of each particle to
the Casimir potential can always be written as:
\begin{equation}
V_{p}=(-1)^{2s_{p}+1}n_{p}\frac{-i}{2}\sum_{n=-\infty}^{\infty}\int\frac{d^{3}p}{(2\pi)^{3}}\log\left(-p^{2}R^{2}+\left(m_{p}\frac{H}{\text{v}}\right)^{2}+\left(n+\theta\right)^{2}R^{2}\right),
\end{equation}
where $s_{p}$, $n_{p}$ , $m_{p}$ are the spin, number of degrees
of freedom and mass of the particle and $\theta$ is a function of
the Wilson lines and $z_{p}$. Note that the masses of the particles appear multiplied by the Higgs divided by its vev, $v$.
The only exception is the Higgs field,
whose mass is changed by the non-linear couplings to the $Z^0$ Wilson line and to itself.
Using the same techniques as e.g. \cite{Ponton:2001hq,HS}  we
regularize the integration and the sum using the analytic continuation
of generalized $\zeta-$functions. After regularization, the contribution
to the Casimir energy for each particle reads:
\begin{equation}
	V_{p}=(-1)^{2s_{p}+1}n_{p}\frac{m_{p}^{2}}{8\pi^{4}R^{4}}\sum_{n=1}^{\infty}\frac{K_{2}(2\pi nR m_{p})}{n^{2}}\text{cos}\left(2\pi n\theta\right)\equiv(-1)^{2s_{p}+1}n_p V_{\mathcal{C}}\left[R,m_{p},\theta\right].
\end{equation}
$K_{2}(x)$ decays
exponentially for large $x$, which means that at low energies, the
contribution of very massive particles to the Casimir energy is highly
suppressed. For massless particles with $\theta=0$ we have
\begin{equation}
V_{\mathcal{C}}\left[R,0,0\right]=\frac{1}{720\pi{R}^{6}},\label{eq:segment_periodic}
\end{equation}
while for $\theta=\frac{1}{2}$ we have
\begin{equation}
V_{\mathcal{C}}\left[R,0,\frac{1}{2}\right]=-\frac{7}{8}\frac{1}{720\pi{R}^{6}}.\label{eq:segment_antiperiodic}
\end{equation}
We cannot use this one-loop formula to study the minima of the potential
around $\sim1$ GeV, since perturbation theory breaks down for the
strong interaction. Below the QCD scale, we
consider an effective field theory of pions and kaons.
For completeness, we give here  the formula above the QCD scale:
\begin{align}
V & =-\frac{(2+2)}{720\pi{R}^{6}}-\frac{2}{8\pi^{5}R^{6}}\sum_{n=1}^{\infty}\frac{1}{n^{4}}\biggl\{2+2\cos\left(2\pi nG_{1}\nonumber\right)\\
& \quad+2\cos\left(2\pi n\left(\frac{1}{2}G_{1}+\frac{3}{2}G_{2}\right)\right)+2\cos\left(2\pi n\left(\frac{1}{2}G_{1}-\frac{3}{2}G_{2}\right)\right)\nonumber\\
& \quad+4\sum_{lept}V_{{\cal C}}\left[R,m_{l}\left(\frac{H}{\text{v}}\right),\frac{1-z_{l}}{2}+Q_{l}A+\frac{1}{2}\left(g_{L}^{l}+g_{R}^{l}\right)Z\right]\nonumber\\
& \quad+4\sum_{flav}\Biggl\{ V_{{\cal C}}\left[R,m_{f}\left(\frac{H}{\text{v}}\right),\frac{1-z_{f}}{2}+Q_{f}A+\frac{1}{2}\left(g_{L}^{f}+g_{R}^{f}\right)Z+\frac{1}{2}\left(G_{1}+G_{2}\right)\right]\nonumber\\
& \quad\quad\qquad+V_{{\cal C}}\left[R,m_{f}\left(\frac{H}{\text{v}}\right),\frac{1-z_{f}}{2}+Q_{f}A+\frac{1}{2}\left(g_{L}^{f}+g_{R}^{f}\right)Z-\frac{1}{2}\left(G_{1}-G_{2}\right)\right]\nonumber\\
& \quad\quad\qquad+V_{{\cal C}}\left[R,m_{f}\left(\frac{H}{\text{v}}\right),\frac{1-z_{f}}{2}+Q_{f}A+\frac{1}{2}\left(g_{L}^{f}+g_{R}^{f}\right)Z-G_{2}\right]\Biggr\}\nonumber\\
& \quad\quad\qquad-6V_{{\cal C}}\left[R,M_{W}\frac{H}{\text{v}},A-c_{w}^{2}Z\right]-3V_{{\cal C}}\left[R,M_{Z}\frac{H}{\text{v}},0\right]\nonumber\\
& \quad\quad\qquad-V_{{\cal C}}\left[R,m_{H}\left(\frac{-1}{2}+\frac{3}{2}\left(\frac{H}{\text{v}}\right)^{2}+\frac{m_{Z}^{2}}{{2}m_{H}^{2}}Z^{2}\right)^{1/2},0\right].\label{eq:tocho}
\end{align}

\begin{figure}
	\centering{}\includegraphics[scale=0.65]{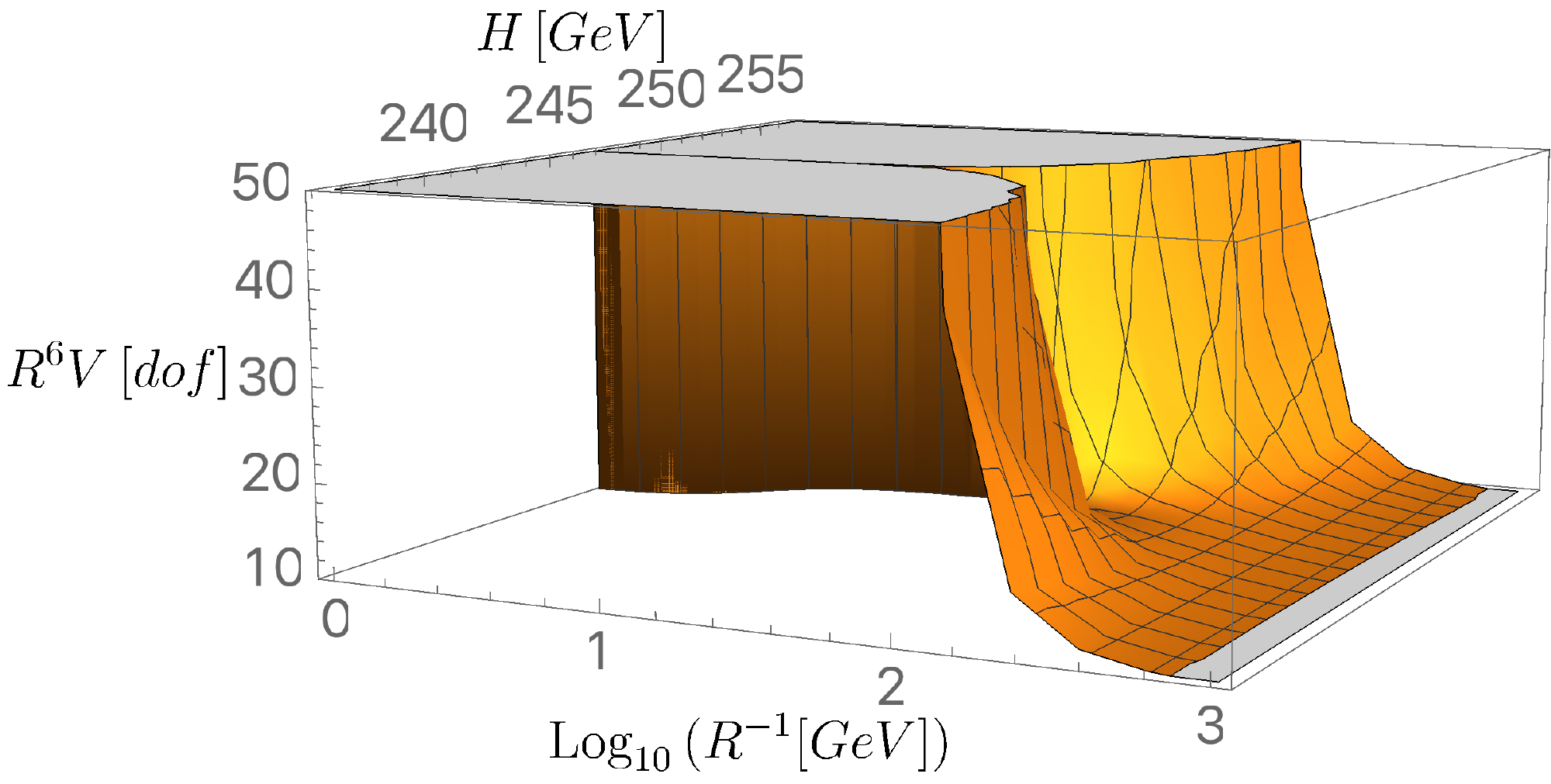}\caption{\footnotesize Effective potential with the Wilson
	lines fixed to zero, as a function of the Radion and the Higgs. The tree level potential
	dominates and the Higgs is not displaced from its tree level minimum
	by the one-loop corrections.  This behavior is independent of the particular value of the Wilson lines
	Although not very visible in the plot, the Higgs minimum remains at the same location as $R^{-1}$ increases.} \label{fig:higgs_at_its_vev}
\end{figure}

The first two lines include the contribution of the graviton, photon and the 8 gluons.  The third line include the one of leptons and the following three lines the 3 colours of
quarks. The seventh line gives the contribution of massive $W^{\pm}$ 's and the $Z^0$. The last line is the contribution of the Higgs. Here $g_{L,R}$ give the 
left and right $Z^0$  couplings  to  fermions, $c_w=\cos \theta_W$ and $Z,A$ are the $Z^0$ and photon Wilson lines. This formula is the generalization of
the one in Appendix B1 in \cite{ArkaniHamed:2007gg}, since it includes
 the $Z^0$ Wilson line  and the Higgs field. Notice that
our gluon Wilson lines are expressed in terms of the basis chosen
in \cite{ArkaniHamed:2007gg}  as
\begin{equation}
	\begin{array}{c}
		G_{1}=G_{\phi}^{1}-G_{\phi}^{2},\\
		G_{2}=G_{\phi}^{1}+G_{\phi}^{2}.
	\end{array}
\end{equation}
Eq. \eqref{eq:tocho} in principle allows  for a detailed study of the SM action on the circle.
In practice,  it is a complicated function of six scalar
fields and the fermions boundary conditions, and a full analysis is a formidable task. 
However a number of conclusions may already be drawn without fully analyzing the complete potential.
 In Fig. \ref{fig:higgs_at_its_vev} we plot the value of  $R^6V$  in the Higgs-Radion plane with Wilson lines
 fixed to zero and periodic boundary conditions for fermions.
 As expected the minima stays  always at $H=\text{v}$. Moreover, this conclusion also holds for different values of the Wilson lines. Thus, we will set the Higgs
equal to its tree level vev from now on.

\begin{figure}
	\centering{}\includegraphics[scale=0.65]{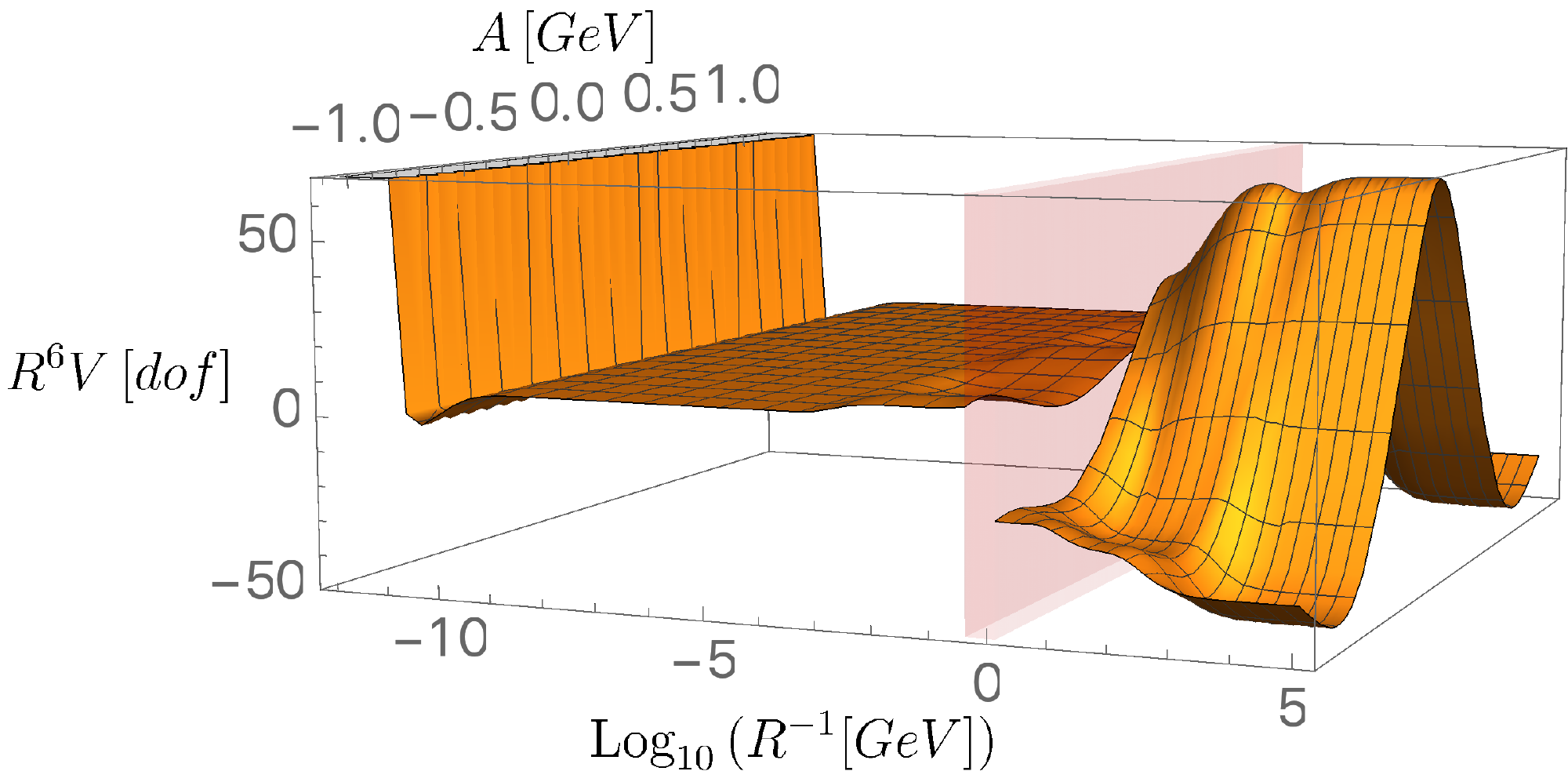}\caption{\footnotesize Effective potential as a function of the photon Wilson line and the
		Radion, with the Higgs field at the minima and all other Wilson lines
		set to zero. Notice that a runaway develops along the Radion direction for a non-zero value of the Wilson line. \label{fig:photon_wl}}
\end{figure}

 Even tough the potential
in terms of the four Wilson lines is still complicated, we can see
with a simple example that the neutrino vacua may become unstable to tunneling
due to the existence of runaway directions at smaller radius R.
In Fig. \ref{fig:photon_wl}
we show the effective potential as a function of the photon Wilson
line and the Radion, with the Higgs field at the minima and all other
Wilson lines set to zero. The point illustrated by this plot is that
the effect of the Wilson lines on the Casimir potential
of a particle is to gradually change its sign. More precisely, it
transforms periodic particles into antiperiodic ones (there is also
a factor $\frac{7}{8}$ involved). The results of  \cite{IMV1}
correspond to the $A=0$, $R^{-1}<m_{e}$ part of the plot,
where we still see the neutrino minima. Moving on to higher energies
along $A=0$, we see a small bump corresponding to the four degrees
of freedom of the electron. After going through the QCD transition,
where we do not know the shape of the potential, we observe a fast 
increase caused by the copiously unleashed quark degrees of freedom.
At  around
$100$ GeV, the $W,Z^0$ and the Higgs slow down a bit the potential
and, finally, the top quark increases it a bit more. Unless new physics
is introduced, $R^{6}V$ remains constant for higher energies,
so $V$ keeps increasing, at least while our one-loop approximation remains valid. On the other hand, if we displace the Wilson line $A$ from
$0$, these conclusions are changed. In particular, for $A=\frac{1}{2}$ or $\frac{3}{4}$, the sign of each contribution is reversed. 
As we can see in Fig. \ref{fig:photon_wl} this leads inevitably to a negative potential. After the top quark scale, $R^{6}V$ remains
constant for higher energies, so $V$ keeps decreasing. This runaway
behaviour of the potential could mean that the neutrino minima is actually metastable, as pointed out in \cite{HS}.  
In fact, to be sure that this vacuum is unstable we should be able to find the bounce interpolating from the vacuum to the runaway direction,
which is a complicated question in such a complicated potential. We will however conservatively assume that indeed the 
vacua are metastable. This would be consistent with the OV conjecture and would lead to
no constraints on neutrino masses nor on the hierarchy.

If we want to search for constraints on observable physics using the OV conjecture we would need  {\it Type P} SM vacua 
in which this potential tunneling in the Wilson line directions is absent.  That would be the case of vacua in which the
Wilson lines are frozen or projected out in such way that these decay directions disappear. In the rest of this paper
we will focus on this class of models. For reasons soon to be explained, a good option is the compactification on the
orbifold $T^ 2/Z_4$. 
But before going to that case we explain in the next section the case of 3D vacua with Wilson line moduli fixed,
which will give us intuition for the more elaborate $T^2/Z_4$ case.

\subsection{Fixing the Wilson lines. The segment $S^1/Z_2$} 
\label{segment}

\begin{figure}
     \begin{center}
     	\subfigure[]{	
        \includegraphics[scale=0.38]{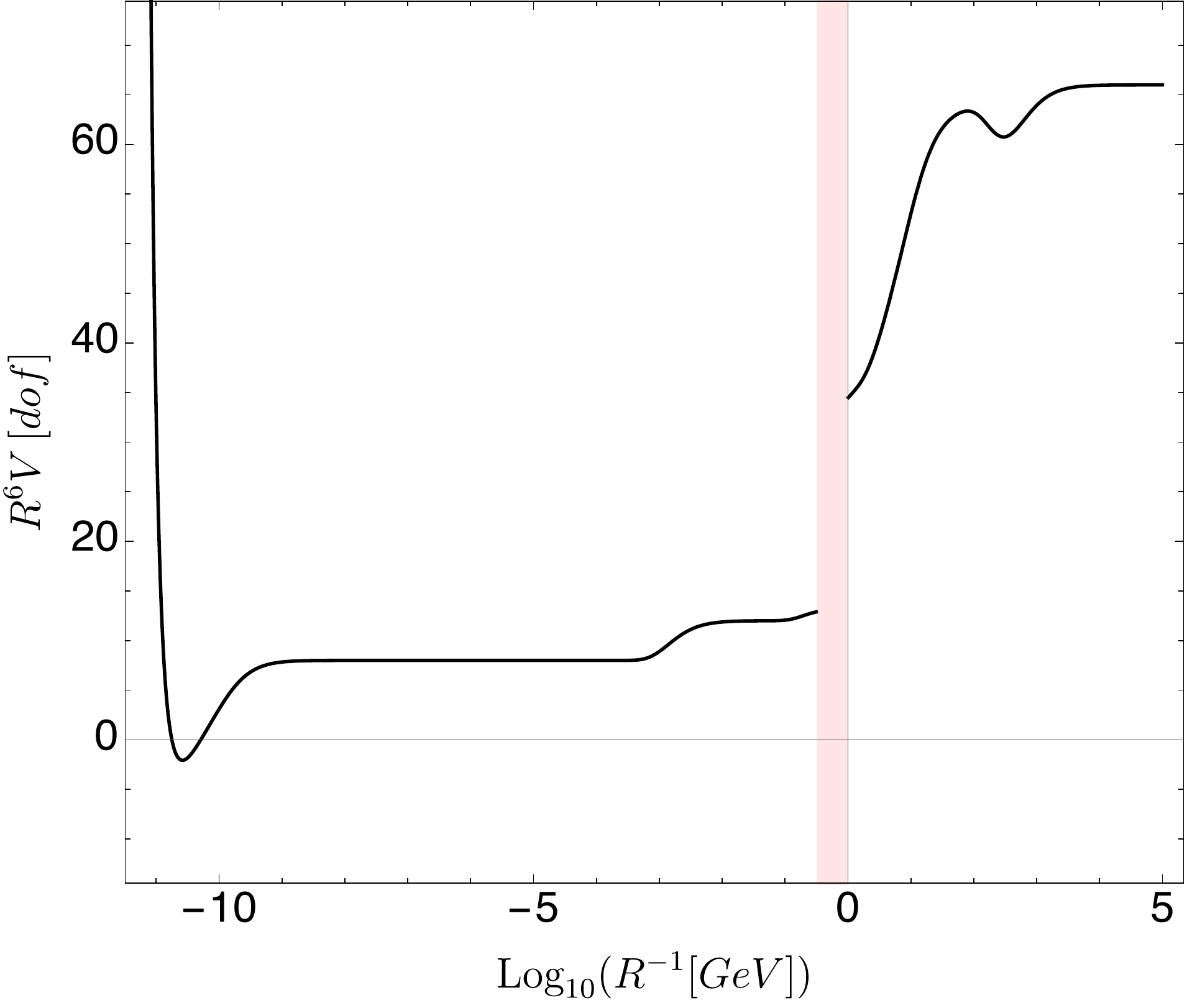} 
       \label{segment_periodic_SMleft}
        }
        \subfigure[]{
        \includegraphics[scale=0.38]{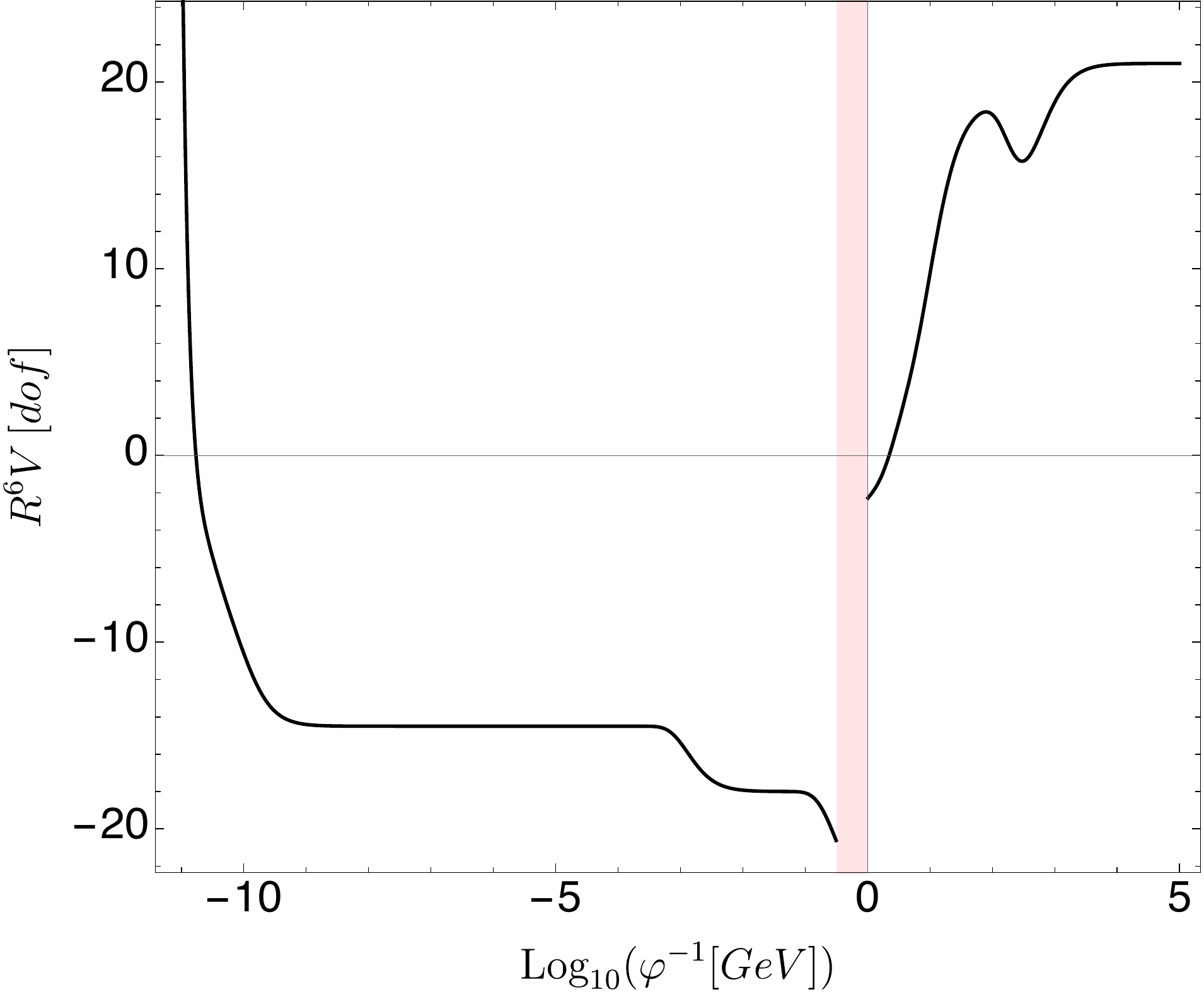}
      \label{segment_periodic_SMright}
        }
      \caption{\protect \footnotesize
		\textbf{(a)}	Effective potential of the Standard Model compactified in a circle with periodic boundary conditions for all particles, the Higgs fixed at its minimum and all Wilson lines turned off. There is no runaway solution. The only AdS minimum we have is the one that would be associated with neutrino bounds. \textbf{(b)} Effective potential of the Standard Model with periodic boundary conditions for all particles except for the leptons. The Higgs is fixed at its minimum and all Wilson lines are turned off. There is no runaway solution. We find an AdS minimum which cannot be avoided by constraining any free parameters of the Standard Model}
      \label{segment_periodic_SM}
     \end{center}
\end{figure}

In this section we describe the structure of the radius scalar potential at shorter distances, up to the EW scale,
setting the Wilson line moduli to zero by hand, looking for new features beyond the neutrino local minima.
In Fig. \ref{segment_periodic_SMleft} we show the product $R^6V$ for 
 the Standard Model with all fermions having periodic boundary
conditions, using the formulae introduced in the previous section for the circle. 
We see that the only minimum we may obtain is the one associated with the
bounds in the neutrino masses (in the Majorana case). Above the
neutrino scale the potential grows smoothly until it reaches the QCD region, around one GeV, where
non-perturbative physics become important and the one-loop Casimir computation is not a good description
(this is denoted by the red, vertical band in the plots). Above the QCD scale perturbation theory again makes sense and the
potential keeps growing indefinitely. The small well in the upper part of the figure corresponds to the  $W^{\pm},Z^0,H^0$ threshold,
while the final increase corresponds to the top quark threshold.
Consequently, the neutrino minima seems stable and the bounds on neutrino masses would apply.

Let us consider now the case in which some of the fermions are instead assigned anti-periodic boundary conditions.
In principle one could naively say that we can assign arbitrary boundary conditions to each fermion 
multiplet of the SM and each  generation. However,
due to the presence of Yukawa couplings and generation mixing all quarks must have the same boundary conditions, and likewise for the leptons. So there are four options according to the 
$(quark,lepton)$ boundary conditions: $(P,P)$, $(AP,P)$, $(P,AP)$ and $(AP,AP)$. The last possibility, with
all fermions antiperiodic does not produce any locally stable vacuum since the potential always decreases and even if they appeared they seem unstable to decay into a {\it bubble of nothing}  \cite{witten}.
It is easy to see that the  $(AP,P)$ case will lead to a runaway potential. In the SM there are more degrees of freedom associated with quarks than leptons and, since for small radius the 
antiperiodic fermion contribution picks up a factor $-\frac{7}{8}$, the potential will grow large and negative.
The remaining case with antiperiodic leptons and periodic quarks is illustrated in 
 Fig. \ref{segment_periodic_SMright}.   As $R$ decreases the potential decreases steadily until the QCD transition. Once passed the hadron region, 
 the potential grows up monotonously due the appearance of  all quark  degrees of freedom, which dominate the
 potential as $R\rightarrow 0$. Although the proximity of the QCD deconfining region does not allow for  a computation of the precise location  of the
 minimum,  one certainly expects a minimum to develop. 
 No lower minima develops for other regions of $R$ and hence that
 minimum seems stable.
This gives a \textit{Type D} minimum since it cannot be avoided by fixing any free parameter
of the Standard Model. For example, the masses of all the SM particles around the QCD scale are known, so there is no way to eliminate it by varying masses,
as it happens in the neutrino case.
As it is, the SM would be inconsistent with the OV conjecture and would not be 
embeddable into a consistent theory of quantum gravity.   Still,  up to here we have not provided a reason why the Wilson lines should 
be fixed.  Those are fixed in the class of $T^2/Z_N$ vacua considered in the next section, and again this class of AdS vacua
will appear. Moreover, we will see in section \ref{sec:SMtorusorbifold} that there are even more AdS vacua which will confirm the difficulties in embedding the SM
into quantum gravity.

A way to avoid the AdS vacua above  would be to go beyond the minimal SM and add e.g. additional massive bosons above the EW scale 
so that the potential develops a runaway behavior as $R\rightarrow 0$, making the dangerous minima unstable.
Indeed, consider the  limit $mR\rightarrow 0$ of the Casimir potential with no Wilson lines.  To first order in the masses of the
particles one obtains
\beqa
V&=&-\frac{(2+2+2\times8)}{720\pi R^{6}}+\sum_{P}(-1)^{2s_{p}+1}\frac{n_{P}}{8\pi}\times\left\{ \frac{1}{90R^{6}}-\frac{m_{p}^{2}}{6R^{4}}\right\}+ 
\nonumber \\ 
&+& \sum_{AP}(-1)^{2s_{p}+1}n_{P}\times\left\{ -\frac{7}{8}\frac{1}{90R^{6}}+\frac{1}{2}\frac{m_{p}^{2}}{6R^{4}}\right\} 
\label{limitesegment}
\eeqa
where we denote by $P$ the particles
with periodic boundary conditions and by AP  the fermions  with
antiperiodic boundary conditions. The first term corresponds to the massless graviton, photon and gluons.
It is clear that if we add a sufficiently large number of  periodic bosons the potential will become negative and
unbounded from below as $R\rightarrow 0$, making the potential unstable, and hence leading to no violation of
the OV conjecture. This scenario is not particularly well motivated. 

We will argue in section \ref{sec:torusMSSMvacua}  that a more elegant option is to embed the SM into a SUSY version like the MSSM. 
It turns out that in the SUSY case this vacuum is not possible since quarks and leptons are forced to have the same boundary conditions. 
But  we will also see that SUSY also avoids new classes of AdS vacua which exist in the orbifold case.
Considering SUSY is an attractive possibility because of another important reason. It is well known that the  Higgs potential of the SM
may have a second  high energy minimum at scales above $10^{10}$ GeV \cite{altominimo,espinosa}. This 4D minimum is would be in  AdS and, if stable, would be again 
inconsistent with the AdS-phobia condition.  Although it would be easy to save the SM by the addition of some BSM physics 
like e.g. additional singlet scalars eliminating the high energy vacuum, SUSY seems to be a more attractive option. If the SUSY breaking scale
is below $10^{10}$ GeV the  high energy Higgs minimum does not develop  (see e.g. \cite{IV}), since the SUSY potential is positive definite. 
We will come back to the SUSY case in section \ref{sec:torusMSSMvacua}.

Up to now we have just set the Wilson lines to zero by hand and have analyzed the resulting potential as a function of 
$R$ and the Higgs field.  We discuss in what follows how the Wilson lines could be fixed.  The simplest option 
seems to be to compactify on the segment $S^1/Z_2$, where $Z_2$ is a reflection with respect to one spatial dimension $y$.
We have boundary conditions under $y\rightarrow -y$  for  scalar, vector and spinor fields:
\begin{align}
	\phi(x_{i},y) & =\phi(x_{i},-y)\\
	A_{\mu}(x_{i},y) & =\left\{ A_{i}(x_{i},-y),-A_{3}(x_{i},-y)\right\}\\
	\Psi(x_{i},y) & =\pm \gamma^{3}\gamma^{5}\Psi(x_{i},-y) .
\end{align}
We see that the Wilson lines are projected out, they are forced to
be zero. Furthermore, half of the Fourier modes are projected out,
and one can check that  the Casimir potential is reduced by a factor 2. 
We can use  directly the
formulas for the circle if we take into account these  two points since
nothing else changes. In this orbifold the contribution to the Casimir
potential of each particle $V_{\mathcal{C}}\left[R,m_{p},\theta\right]$
is thus very  simple.  The results for the scalar potential would be qualitatively identical to the
case of the circle with Wilson lines fixed to zero.

We could envisage to compactify the SM on the segment and obtain the results above with fixed Wilson lines. 
However  there is a technical obstruction 
in this simple situation.  Indeed, reflection with respect to a single  coordinate (or in general, parity) is not a symmetry of
the SM  (since it reverses chirality) and we cannot twist the theory with respect to it, unless we do some modification or extension. 
Such an extension could be the embedding of the SM in a theory which is left-right symmetric like
$SU(3)\times SU(2)_L\times SU(2)_R\times U(1)$, $SU(4)\times SU(2)_L\times SU(2)_R$ or $SO(10)$.
But then we have to include the Higgsing down to the SM in the computations. We will prefer instead to
go to 2D orbifold compactifications $T^2/Z_N$, with $Z_N$ a discrete rotation in two dimensions,
which is always a symmetry of the 4D SM. We will first discuss the case of the torus, from which it is easy to obtain
the results for the orbifolds.

\section{The SM on the  $T^2/Z_4$ orbifold}
\label{sec:SMtorusorbifold}

In this section we consider the case of the compactification of the SM on the torus $T^2$ and then on the orbifold $T^2/Z_4$ which is
closely related. 

\subsection{The SM on the $T^2$ torus} 

In this section we first review some of the most relevant features and results of the compactification of the SM on a torus. 
Compactifications of the SM on a torus have been considered in \cite{ArkaniHamed:2007gg, Arnold:2010qz, Arnold:2011cz} and \cite{ IMV1, HS}. 

As in the circle, to obtain the potential of the moduli
we will consider the  scalar potential of the SM plus Einstein gravity compactified on the torus and the one-loop contribution coming from the Casimir energy of the corresponding particles.

We take for the distance element in 4 dimensions,
\begin{equation}
ds^2= g_{\alpha \beta}  dx^\alpha dx^\beta+ B_{\alpha i} dx^\alpha dy^i +t_{ij} dy^i dy^j,
\end{equation}
with $g_{\alpha \beta}$ the 2-dimensional metric, $B_{\alpha 2}$ and $B_{\alpha 3}$  the two graviphotons, which will be set to zero from now on for the same reason as in the circle, and $t_{ij}$ the metric on the torus, which reads
\begin{equation}
t_{ij}=\frac{a^2}{\tau_2}\left[\begin{array}{cc}
1&\tau_1 \\
\tau_1& |\tau|^2 
\end{array}\right].
\end{equation}
Here $\tau=\tau_1+i\tau_2$ and $a^2$  are the complex structure and area moduli, respectively. The relevant piece of the effective action after dimensional reduction on the torus takes the form
\begin{equation}
S_{GR+SM}= \int d^2x \sqrt{-g_2} \left[ \dfrac{1}{2}M_p^2  (2  \pi a)^2\left(  \mathcal{R}_{(2)} - \dfrac{1}{2 \tau_2} \left\{ (\partial_\alpha \tau_1)^2+ \partial_\alpha \tau_2)^2\right\}\right) -V \right],
\end{equation}
where $V$ includes the 4D cosmological constant, the tree-level and the Casimir contributions to the potential of the scalars in the theory. Notice that, as expected from the fact that the 4-dimensional graviton has two degrees of freedom, only the two complex structure moduli propagate in two dimensions, whereas the area moduli does not propagate any additional degree of freedom (i.e. it has no kinetic term in the 2D action). Moreover,
unlike in the case of the circle, 
now we cannot perform a Weyl transformation to express the action in the Einstein frame, due to the conformal invariance of 2D gravity. This means that the function $V$ should not be interpreted as a canonical potential, and that we have to resort to the equations of motion in order to study the vacua of the lower dimensional theory. Even though $V$ is not minimized with respect to all the variables at the vacua of the theory, we will still call it  potential. The conditions for AdS$_2$ vacua are (see \cite{ArkaniHamed:2007gg, Arnold:2010qz} for details):
\begin{equation}
\begin{array}{l c l c l}
V(a,\tau, \phi_j)=0, & & \partial_{\tau_1,2}V(a,\tau, \phi_j)=0, & & \partial_{\phi_j} V(a,\tau, \phi_j)=0, \\
\partial_a V(a,\tau, \phi_j)<0, & & \partial^2_{\tau_1, \tau_2, }V(a,\tau, \phi)>0, & & \partial^2_{\phi_j} V(a,\tau, \phi_j)>0,\\
\end{array}
\end{equation}
where the $\phi_j$ represent all the other 2-dimensional scalars in the theory. Notice that these conditions are the usual minimization conditions for the potential with respect to all the scalars in the theory except  $a$. These two atypical constraints (the ones in the first column) come from requiring constant field solutions to the equations of motion.
 Variation of the action with respect to $a$ yields the constraint
\begin{equation}
\mathcal{R}_{(2)}=\dfrac{1}{(2 \pi M_p)^{2}}\dfrac{\partial_a V}{a} \ ,
\end{equation}
so that the curvature in 2D is fixed by the derivative of the potential with respect to $a$.  Thus an AdS vacuum is obtained for a negative derivative.
A possible new degree of freedom in the torus case is the addition of magnetic fluxes along some $U(1)$ in the Cartan subalgebra of the SM. However,
fluxes contribute enormously to the vacuum energy and do not lead to interesting vacua, see \cite{Arnold:2010qz,HS}.

Regarding the rest of 2D scalars that enter the potential, we have the Higgs, two Wilson lines (corresponding to the two cycles of the torus) for each of the neutral gauge fields in the SM, that is, two for the photon $A$, two for the $Z^0$ boson and two more for each of the two gluon fields belonging to the Cartan subalgebra of $SU(3)_C$. The potential that we consider has the following structure
\begin{equation}
V[a, \tau,H,Z_i,A_i,G_{1i},G_{2i} ]  = (2\pi a)^2\Lambda_{4} + (2\pi a)^2V_{\mathrm{SM}} (H, Z_1, Z_2) + \sum_{p}V_{p}(a, \tau,H,Z_i,A_i,G_{1i},G_{2i}).
\end{equation}
As in the circle compactification, it consists of a piece that depends on the 4D cosmological constant $\Lambda_4$, a tree level piece for the Higgs and the $Z$ Wilson lines (which comes from the 4D Higgs potential and the 4D  mass term of the $Z$ boson, respectively) and a contribution from the Casimir energy of the particles in the spectrum. Again, this last piece is the sum of the contributions coming from each particle and introduces the dependence on the geometric moduli, the Higgs and the Wilson lines to which it couples. In order to compute the Casimir energy we need to expand our 4D fields in a basis that respects the periodic ($z_i=1$) or antiperiodic ($z_i=0$) boundary conditions along the two cycles of the torus, that is
\begin{equation}
\phi(x^\alpha, y^i) =\sum_{n_{1},n_{2}=-\infty}^\infty \phi_{n_{1}, n_{2}}(x^\alpha)e^{2\pi i \left[ \left(n_1+\frac{1-z_1}{2} \right)y_1 +\left(n_2+\frac{1-z_2}{2} \right)y_2 \right]}.
\end{equation}
After inserting this expansion into the one-loop quantum effective action and integrating over  the compact dimensions, the Casimir contribution of a particle with spin $s_p$, $n_p$ degrees of freedom and mass $m_p$ turns out to be:
\begin{equation}
V_p= (-1)^{2s_p+1}n_p\frac{-i}{2}\sum_{n_1,n_2=-\infty}^\infty \int \dfrac{d^{2}p}{(2 \pi)^2} \log \left[ -p^2 + \left(m_{p} \dfrac{H}{v} \right)^2+\dfrac{1}{a^2 \tau_2}\left|  \left( n_2 + \theta_2 \right) - \tau \left( n_1 + \theta_1 \right)\right| ^2 \right].\label{torusintegral}
\end{equation}
The $\theta_{i}$ are functions of the appropriate Wilson lines and boundary conditions $z_i$. Using dimensional or $\zeta$-function regularization techniques one arrives, after renormalizing the cosmological constant,  at the following expression for the Casimir energy of each particle \cite{HS, Arnold:2010qz}:
\begin{align}
V_p&= \dfrac{(-1)^{2s_p+1} n_p}{(2\pi a)^2}\left[2 a^{2} \tau_2 m^2  \sum_{p=1}^\infty \frac{\cos (2\pi p \theta_1)}{p^{2}}K_2 \left(\dfrac{2 \pi p a m}{\sqrt{\tau_2}} \right)+ \right.\nonumber \\
&\left. +\dfrac{1}{4 \pi \tau_2} \sum_{n=-\infty}^{\infty}   \left(  2 \pi \sqrt{(n+\theta_1)^2\tau_2^2+m^2 a^2 \tau_2 }  \left\{ \mathrm{Li}_2\left(e^{\sigma_+}\right)+\mathrm{Li}_2(e^{\sigma_-})\right\}  +  \left\{ \mathrm{Li}_3(e^{\sigma_+}) +\mathrm{Li}_3(e^{\sigma_-})\right\} \right)\right] \nonumber\\
&\equiv (-1)^{2s_p+1} n_p V_{\cal{C}}[a,\tau,m,\theta_1, \theta_2], \label{toruscasimir}
\end{align}
where $\mathrm{Li}_s$ is a polylogarithm and 
\begin{equation}
\sigma_{\pm}=2\pi\left(\pm i\left\{ -\left(n+\theta_{1}\right)\tau_{1}+\theta_{2}\right\} -\sqrt{\left(n+\theta_{1}\right)^{2}\tau_{2}^{2}+Ma^{2}\tau_{2}}\right)\label{eq:sigmas}.
\end{equation}
There are five scalar fields that must be stabilized at low energies (the three geometric moduli and the two photon Wilson lines), and the number increases  as the energy grows: four extra gluon Wilson lines once the QCD scale is surpassed and the Higgs plus the two $Z^0$ Wilson lines when the EW scale is reached. In the case of the circle we saw that the Wilson lines tend to create runaway directions that non-perturbatively destabilize  any  possible AdS vacua 
 of the lower dimensional theory \cite{HS}, hence ruining the possibility to obtain any constraint from AdS-phobia,
we only obtain {\it Type S} vacua. As we saw in subsection \ref{segment}, setting the Wilson lines to zero simplifies things in such a way that constraints are expected 
 to come up. This is our main motivation to consider an orbifold compactification of the torus, as the action of the $Z_N$ will project out the Wilson lines from the spectrum. 

Let us explain why we  choose to study, in particular, the $Z_{4}$ orbifold. First of all the complex structure is only fixed if the quotient is by $Z_N$ with $N \geq 3$. The quotient by $Z_2$ would project out the Wilson lines but would not fix the complex structure. In this case it is necessary to find the minima in the direction of the complex structure. This is easy to do when all particles have periodic boundary conditions. In this case the potential is invariant under $SL(2;Z)$ modular transformations and therefore the extrema of the potential must be at the stationary points of the complex structure, that is, $\tau=1$ or $\tau=1/2 + i \sqrt{3}/2$. In fact, as it is concluded in \cite{Arnold:2010qz}, only the latter allows for the existence of minima.
However, if there are both P and AP boundary conditions the combined minimization of complex structure and radius becomes less trivial.
  As shown in subsection \ref{segment}, the appearance of antiperiodic boundary conditions is of some interest, because they may generate additional ``exotic''AdS minima, that come out as a generic feature of the SM as we know it. For this reason, we prefer an orbifold in which the complex structure is fixed from the start.
   Finally, among the remaining orbifolds,  the formulas for the $Z_{4}$ orbifold are the easiest ones to work with. Still the general results obtained in the next section are expected to apply for any 
$Z_N$ orbifold acting crystallographically on the torus.

\subsection{SM compactification  on the  $T^{2}/Z_{4}$ orbifold}

The most relevant difference between $T^{2}$ and $T^{2}/Z_{4}$ is that we need to introduce additional boundary conditions
to prevent the Lagrangian from becoming multivalued as a result of
the new identifications. Instead of using $a,\tau$ it is sometimes useful
to use the torus radii $R_{1},R_{2}$ and the angle $\theta$ between
them as parameters:
\begin{equation}
\tau=e^{i\beta}\frac{R_{2}}{R_{1}}\qquad \tau_{2}=\frac{a^{2}}{R_{1}^{2}}\qquad \tau_{1}=\frac{1}{R_{1}^{2}}\sqrt{R_{1}^{2}R_{2}^{2}-a^{4}}\qquad a^{2}=R_{1}R_{2}\sin\beta
\end{equation}
Remember that, for consistency of the $Z_{4}$ identifications \cite{quevedo}, the geometrical action of the orbifold must act crystallographically on the torus lattice, fixing the complex structure to $\tau=i$ and $R_{1}=R_{2}\equiv R$ or $a^{2}=R^{2}$. We have
two symmetry transformations associated with the torus ${\cal T}$
and three non-trivial ones with the orbifold ${\cal P}$. Their action on the compact
coordinates is given by:
\begin{align}
{\cal T}_{A}(y_{1},y_{2}) & =(y_{1}+R,y_{2})\\
{\cal T}_{B}(y_{1},y_{2}) & =(y_{1},y_{2}+R)\\
{\cal P}_{4}(y_{1},y_{2}) & =(-y_{2},y_{1})\label{eq:orbione}\\
{\cal P}_{4}^{2}(y_{1},y_{2}) & =(-y_{1},-y_{2})\label{eq:orbitwo}\\
{\cal P}_{4}^{3}(y_{1},y_{2}) & =(y_{2},-y_{1})\label{eq:orbithree}
\end{align}
Under any of these symmetries the scalar fields
transform as $\phi\left(x,y\right)\longrightarrow\phi\left(x,y'\right)$. It is enough to impose boundary conditions for the two translations and for the smallest rotation, of angle $\frac{\pi}{2}$. The boundary conditions for the translations imply that we can expand the modes in the usual Fourier series:
\begin{align}
\phi\left(x_{i},y_{1},y_{2}\right) & =\sum_{n_{1},n_{2}=0}^{\infty}\biggl\{\phi_{n_{1}n_{2}}^{++}\left(x_{i}\right)\cos2\pi n_{1}y_{1}\cos2\pi n_{2}y_{2}+\phi_{n_{1}n_{2}}^{+-}\left(x_{i}\right)\cos2\pi n_{1}y_{1}\sin2\pi n_{2}y_{2}\nonumber \\
& \qquad\qquad+\phi_{n_{1}n_{2}}^{-+}\left(x_{i}\right)\sin2\pi n_{1}y_{1}\cos2\pi n_{2}y_{2}+\phi_{n_{1}n_{2}}^{--}\left(x_{i}\right)\sin2\pi n_{1}y_{1}\sin2\pi n_{2}y_{2}\biggl\}\label{eq:scalar_expansion}
\end{align}
We will be particularly interested in Lagrangians which are invariant under some discrete gauge group, possibly embedded at some scale in a gauged $U(1)$.
In this case we can complement the geometrical action with an internal one belonging to the discrete gauge subgroup. 
Thus e.g., in the case of a $U(1)$ symmetry and denoting their charge scalar by $q$, the boundary condition for the $\frac{\pi}{2}$ rotation is given by
   \begin{equation}
  \phi(x_{i},y_{1},y_{2})=e^{i q \alpha} \phi(x_{i},-y_{2},y_{1}) \ ,
  \label{scalarboundary}
  \end{equation}
  for some real $\alpha$ which is the same for all of the particles involved in the $U(1)$ symmetry. It is then easy to see that a solution exists for $q\alpha= 2k \frac{\pi}{4},$  where $k=0,\pm 1, \pm 2...$ Without loss of generality we fix $\alpha=\frac{\pi}{4}$ so that the charges are just even integers $q=2k$ for scalars.
   In particular, for even $k$ the mode is given by: $\phi_{n_{1}n_{2}}^{++}=\pm \phi_{n_{2}n_{1}}^{++}$, $\phi_{n_{1}n_{2}}^{--}= \mp \phi_{n_{2}n_{1}}^{--}$, $\phi_{n_{1}n_{2}}^{+-}= \phi_{n_{1}n_{2}}^{-+}=0$. For odd $k$ it is given by $\phi_{n_{1}n_{2}}^{+-}=\pm i \phi_{n_{2}n_{1}}^{-+}$,  $\phi_{n_{1}n_{2}}^{++}= \phi_{n_{1}n_{2}}^{--}=0$. In all of these cases the linearly independent Fourier modes can be confined to the positive $n_{1},n_{2} \geq 0$ sector. Notice that the zero mode survives only for $k$ multiple of $4$.

Regarding the fermions we also start with the boundary conditions for the torus translations. Periodic or antiperiodic boundary conditions are taken into account by replacing $n$ with $\tilde{n}=n+\frac{1-z_{p}}{2}$ in the usual Fourier series. Under a general rotation Weyl spinors transform as $\psi\left(x,y\right)\longrightarrow e^{i\frac{\vec{\varphi}}{2}\vec{\sigma}}\psi\left(x,y'\right)$. Taking the $\frac{\pi}{2}$ rotation to be around the third axis we find: \begin{equation}
\psi\left(x_{i},y_{1},y_{2}\right)\longrightarrow 
\left(\begin{array}{cc}
e^{i\frac{\pi}{4}} & 0\\
0 & e^{-i\frac{\pi}{4}}
\end{array}\right)\psi\left(x,y'\right).
\end{equation}
If the internal global symmetry is a U(1), then the boundary condition for the orbifold rotation Eq. \eqref{eq:orbione} is:
\begin{equation}
\psi\left(x_{i},y_{1},y_{2}\right))=e^{i q \alpha} e^{i\frac{\pi}{4}\sigma_{3}}\psi\left(x,-y_{2},y_{1}\right).
\label{fermionboundary}
\end{equation}
One finds that for $\alpha=1/4$ a solution exists only for odd $q$. Besides, it is always possible to confine all the independent Fourier modes to the positive $n_{1},n_{2} \geq 0$ sector.
For $q=1+8k$ we obtain a 1-component 2D  fermion in the massless sector, otherwise there is no zero mode for the fermion.
To sum up, depending on its charge, each particle may or may not have a zero mode and also a non-vanishing contribution to the Casimir energy (i.e. it contributes only if it admits an expansion, that is, it has a KK tower). Let us recall that only massive KK excitations contribute to the Casimir potential, the possible zero modes do not. Hence the crucial feature in order to
compute the potential is the presence or not of a KK tower for a given particle.

Finally, vector fields transform as $PA_{\mu}(x^{\sigma})P^{-1}={\cal P}_{\mu}^{\:\:\:\nu}A_{\nu}({\cal P}_{\: \: \: \: \rho}^{-1\:\: \sigma}x^{\rho})$. Since they transform in the adjoint of their gauge group the orbifold action is given by:
\begin{equation}
A_{\mu}\left(x_{i},y_{1},y_{2}\right)=U\left\{ A_{i}\left(x_{i},-y_{2},y_{1}\right),-A_{3}\left(x_{i},-y_{2},y_{1}\right),A_{2}\left(x_{i},-y_{2},y_{1}\right)\right\}U^{\dagger}.
\end{equation}
In the case of a U(1) symmetry, the U's cancel out. The key observation is that the boundary conditions project out the Wilson lines. As explained in the text, this is the main reason why we study orbifolds. Again, the most general solution is a combination of positive modes. Note that one can add however quantized Wilson line backgrounds of order 4.  We have checked that this extra discrete degree of freedom 
only gives rise to new {\it Type S} vacua which are of no relevance for the present discussion.

Once we have found the expansion for each field, we insert it in the path integral, through
the action, and proceed in the same way as in the torus or the circle. After integration in the compact coordinates, the only difference between $T^{2}$ and $T^{2}/Z_{4}$ is in the values that $\tilde{n}_{1},\tilde{n}_{2}$ can take. We have seen that in $T^{2}/Z_{4}$ the independent normal modes can be confined to the positive $\tilde{n}_{1},\tilde{n}_{2}\geq 0$ sector. Thus, the potential for the orbifold is given by the torus result 
 Eq. \eqref{torusintegral} evaluated at $\tau=i$ and $\theta_{1}=\theta_{2}=\theta$ and with the restriction $\tilde{n}_{1},\tilde{n}_{2}\geq 0$:
\begin{equation}
V_p= (-1)^{2s_p+1}n_p\frac{-i}{2}\sum_{n_{1},n_{2}=0}^\infty \int \dfrac{d^{2}p}{(2 \pi)^2} \log \left[ -p^2 + \left(m_{p} \dfrac{H}{v} \right)^2+\dfrac{1}{a^2}\left(\tilde{n}^{2}_2 + \tilde{n}^{2}_1 \right) \right].
\end{equation}
Now, because we are at $\tau=i$, the modified indices $\tilde{n}_{1}, \tilde{n}_{2}$ appear only squared. This means that we can extend the summation in $n_{1},n_{2}$ to the integers (remember that the zero mode, $n_1= n_2 =0$,  does not contribute to the Casimir energy and that $\tilde{n}=n+\theta$) as in the general formula of the torus. For this reason, we could just use the general formula, Eq. \eqref{toruscasimir} evaluated at $\tau=i$ and with an extra factor $\frac{1}{4}$. 

For $\theta=0$ it is shown in  Appendix \ref{app:1looptorusorbifold} that the contribution from a massless particle can be expressed in a simple form:
\begin{equation}
V_{\mathcal{C}}\left[a,\tau=i,0,0,0\right]=\dfrac{1}{(2\pi a)^2}\dfrac{\mathcal{G}}{3},\label{orbifoldperiodic}
\end{equation}
where 
\beq
\mathcal{G}\equiv \sum_{n=0}^{\infty}{\frac{(-1)^{n}}{(2n+1)^{2}}}
\eeq
is Catalan's constant, $\mathcal{G}\simeq 0.915966$.
 For $\theta=\frac{1}{2}$, the Casimir energy reads
\begin{equation}
V_{\mathcal{C}}\left[a,\tau=i,0,\dfrac{1}{2},\dfrac{1}{2}\right]=\dfrac{-1}{(2\pi a)^2}\dfrac{\mathcal{G}}{6}=-\dfrac{1}{2} V_{\mathcal{C}}\left[a,\tau=i,0,0,0\right].\label{orbifoldantiperiodic}
\end{equation}

The discussion for the massive case is essentially the same as for the segment. For $a^{-1}\ll m_{p}$ (the particle mass), the potential is exponentially suppressed with
the mass, while for $a^{-1}\gg m_{p}$ it tends to $V_{\mathcal{C}}\left[a,\tau=i,0,\theta,\theta\right]$, growing as $\propto a^{-2}$. 
In the case of  $S^1/Z_2$ the potential for a massless particle decayed as $\propto R^{-6}$ and that is why $R^{6}V$ was constant in the massless limit and easier to plot.  We had to look for the minima followed by zeros of $R^{6}V$, since only those meant a minima for $V$. Now we are concerned with the zeros of $a^{2}V$ as a function of $a$, since those are also zeros of $V$. In both cases we can normalize the potential in units of degrees of freedom. Finally, as in the case of the segment, fermions can be either periodic or antiperiodc. The only difference is that the factor $-\frac{7}{8}$ is now replaced by a factor $-\frac{1}{2}$.

\subsection{Embedding $Z_4$ into discrete gauge symmetries}

From the discussion above, particularly Eq.\eqref{fermionboundary}, one observes that if fermions are not charged under a  discrete 
subgroup of a $U(1)$ symmetry (i.e. $\alpha=0$) there is no KK tower for any fermions in the theory. Then the KK towers are purely 
bosonic and the Casimir potential becomes unbounded from below as $a^2\rightarrow 0$. This means we get a vacuum of {\it Type S} which
is consistent with AdS-phobia but leads to no constraint. In order to obtain  {\it Type P} vacua leading to interesting constraints on the SM we need to
embed the $Z_4$ action into  an order eight discrete subgroup of a gauge (or gaugable) $U(1)$ symmetry, to match the $e^{i\pi/4}$ phase 
that fermions get under a $\pi/2$ rotation. In this way  the SM fermions 
may get KK excitations and contribute to the Casimir energy, which is what we are searching for in order for neutrinos to play a role.
We will consider three possibilities for such an embedding
\begin{itemize}
\item {\it a)}  Discrete subgroups of the Cartan subalgebra of the SM gauge group $SU(3)\times SU(2)\times U(1)$.
\item {\it b)} Discrete gauge symmetries. We will consider symmetries which are subgroups of $U(1)$ global symmetries of the SM and may be gauged at a higher scale.
This  is the case of $(B-L)$ which is 
a global anomaly-free symmetry of the SM (if right-handed neutrinos are present) and 
is also a global gauge symmetry of the R-parity conserving MSSM. If the  SM or MSSM is extended by a $U(1)_{B-L}$ gauge symmetry,  this embedding will 
always exist. However 
discrete $Z_N$ subgroups of $B-L$ may be gauged with no need for the 
full gauge group $U(1)_{B-L}$ to be gauged at any scale, see e.g.\cite{discretas}.

\item {\it c)} Discrete R-symmetries in the case of the SUSY SM.  Those may appear in string compactifications as gauged R-symmetries,
see e.g.\cite{discretestring}.
\end{itemize}
The embeddings of the first type are always available and hence the conclusions one derives from the existence or not of AdS vacua 
in that  case will always apply.  For the other symmetries the conclusions will depend on whether the theory underlying the 4D string compactification
leading to the SM admits those symmetries. In particular if $U(1)_{B-L}$ is gauged, we will always be allowed to twist by any discrete subgroup of it.

For convenience of the reader, in Table \ref{cargas} we show the quantum numbers of SM or MSSM particles with respect to some relevant 
$U(1)$ symmetries and R-symmetries (in the MSSM case).  We will consider twisting with respect to discrete subgroups of them, as well
as the Cartan subalgebra of $SU(3)\times SU(2)$.
\begin{table}[htb] \footnotesize
\renewcommand{\arraystretch}{1.25}
\begin{center}
\begin{tabular}{|c||c|c|c|c|c|c||c|c||c|}
\hline  Generator 
                      &  Q &  u  &   d & L & e & $\nu_R$ & $\tilde{H}$ & $\tilde{{\bar H}}$  & ${\tilde g},{\tilde W},{\tilde B}$ \\
\hline
   Y                &   1   & -4 & 2  & -3 & 6 & 0 & -3 & 3 &   0 \\
   \hline
   B-L                &   1   & -1  & -1   & -3 &  3 & 3 &  0 & 0 &   0 \\
   \hline
   $U(1)_s$       &   -1   & -1 & -1  & -1 & -1 & -1 &  1 & 1 &   1 \\
\hline
\end{tabular}
\end{center} \caption{ Some fermion $U(1)$ q-charges considered in the text. The line  in the bottom is the R-symmetry of the MSSM and the 
corresponding sfermions have charge $q+1$.}
\label{cargas} 
\end{table}

\subsubsection{AdS SM  $Z_4$ vacua}

It is easy to check that embedding $Z_4$ into a discrete subgroup (of order eight) of weak hypercharge $Y$ or the third component of weak isospin
leads to uninteresting but safe {\it Type S} vacua. Indeed the reader may check that only left-handed quarks and leptons build a KK 
tower and contribute  in this case.  Since the right-handed fermions are projected out, no masses can appear for the 
fermions. Thus the KK towers are dominated by these massless fermions,  which overwhelm the contributions of the photon and graviton  avoiding   AdS minima to develop.

\begin{figure}
	\begin{center}
		\subfigure[]{	
			\includegraphics[scale=0.38]{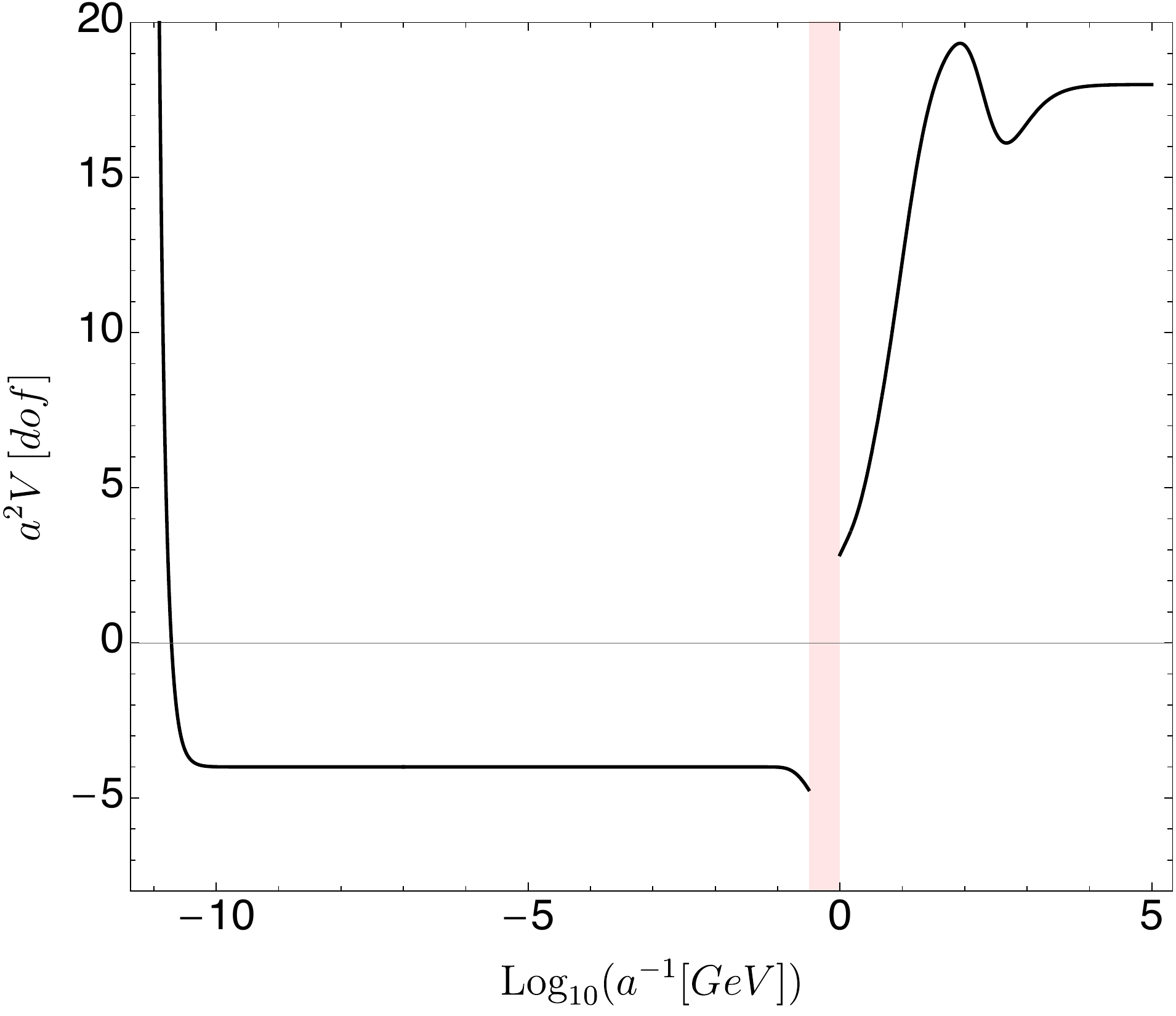} 
			\label{su3SM}
		}
		\subfigure[]{
			\includegraphics[scale=0.38]{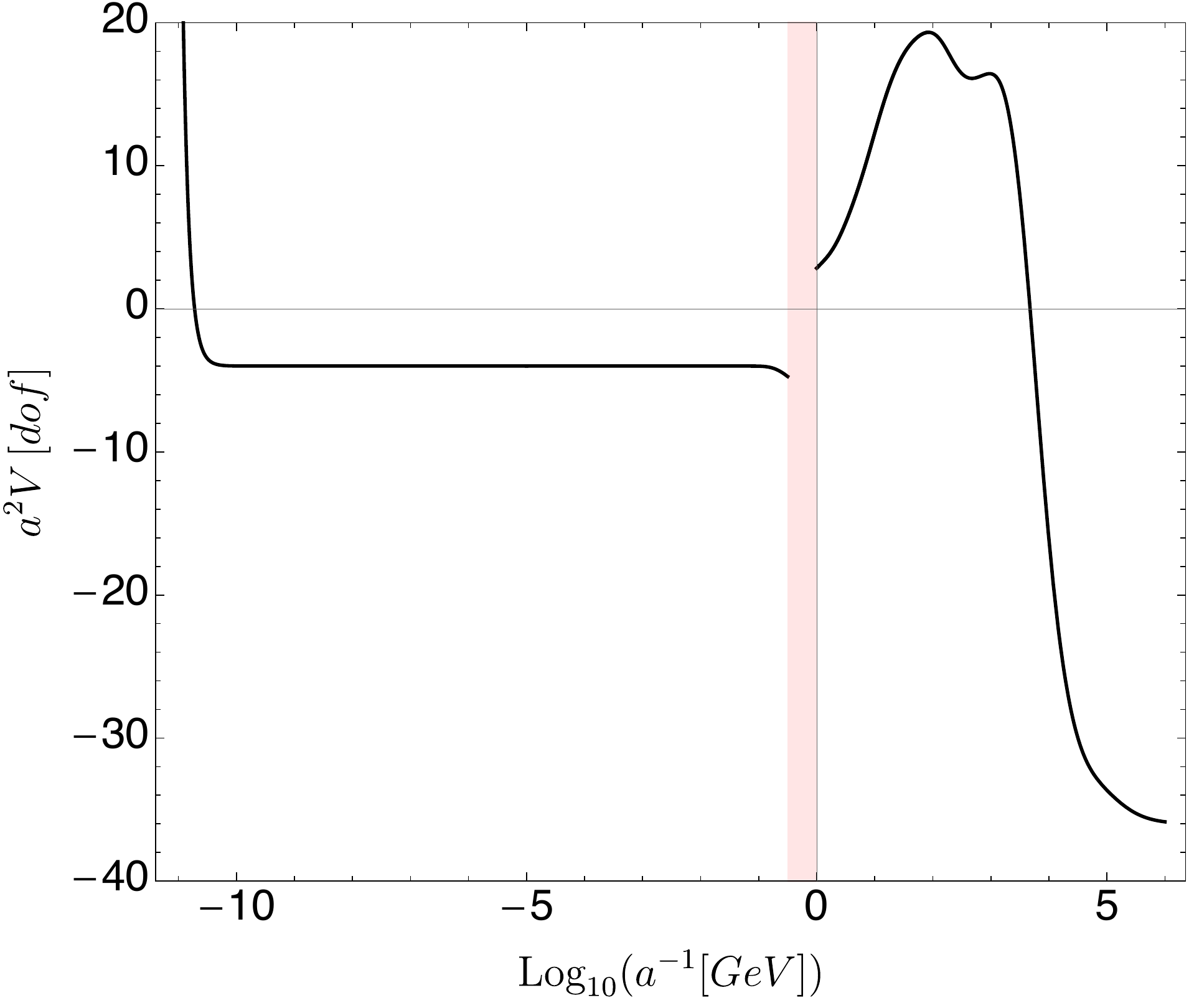}
			\label{su3MSSM}
		}
		\caption{\protect \footnotesize
			\textbf{(a)}	Effective potential of the Standard Model compactified in $T^{2}/Z_{4}$ embedded into a discrete  subgroup of the $SU(3)$ colour Cartan subalgebra. All particles have periodic boundary conditions. As usual, the Higgs is fixed at its minimum. There is no runaway solution, so the minimum which appears around the QCD transition is of type D. The SM as it is appears to be inconsistent with Quantum Gravity. \textbf{(b)} The same plot but for the MSSM. The addition of extra bosonic degrees of freedom creates a runaway behaviour in the potential which could make the minima unstable. The MSSM would then appear to be consistent with the no stable AdS hypothesis. }
		\label{segment_periodic_SM}
	\end{center}
\end{figure}

On the other hand if we embed  $Z_4$ into a discrete  subgroup of the $SU(3)$ colour Cartan subalgebra, {\it Type D} minima develop. To see how this comes about, we begin by assigning periodic boundary conditions for the torus translations to all particles. Regarding the orbifold rotation boundary conditions, notice that all of the particles except for quarks and gluons have zero (strong) charge. From Eqs. (\ref{scalarboundary}) and (\ref{fermionboundary}) this means that the Higgs bosons survive but the leptons are projected out. For the quarks the boundary condition is given by the following equation, diagonal in colour space: 
\begin{equation}
\Psi\left(x_{i},y_{1},y_{2}\right)=e^{i\left(q_{3}T_{3}+q_{8}T_{8}\right) } e^{i\frac{\pi}{4}\sigma_{3}}\psi\left(x,-y_{2},y_{1}\right).  \label{eq:quarkglobal}
\end{equation}
We know that a colour will survive only if its corresponding charge is odd. Introducing the Gell-Mann matrices we 
find these three equations: $q_{3} + q_{8}=odd$, $- q_{3} + q_{8}=odd$ and $2 q_{8}=odd$ for the three colours respectively.
If the first two are satisfied then the third one is not satisfied, as can be seen by adding them.  Thus, only two colours of  quarks will get KK towers and no leptons.
Doing the same for the gluons, one finds that  SU(3) is broken to $SU(2)\times U(1)$.  
With this spectra, the contribution from the photon and graviton 
dominates down  to the QCD transition, where the quarks make the potential positive. Thus, a minimum always develops, see Fig. \ref{su3SM}.
So, remarkably,  the SM as it is would be {\it excluded by the existence of this AdS vacuum}. There is no dependence on any particle mass
(like the lightest neutrino mass in the neutrino type of vacua) which could save the day. As we said, the reason why the SM necessarily
has AdS vacua is that it is too fermionic.  If the theory  had more bosons than fermions this vacuum would become unstable and
consistency with the WGC would be obtained.  That  is precisely what is naturally provided by a SUSY version of the SM
which we consider next.

\section{ $Z_4$  MSSM vacua}
\label{sec:torusMSSMvacua}

We now turn to the case of the MSSM. We consider here for definiteness the case of the simplest R-parity conserving 
version of the SM.  Thus we have  sfermions partners for fermions, Higgsinos for Higgses (we assume a minimal set) and gauge
bosons come with gauginos. 
Let us check first whether twisting by discrete subgroups of the Cartan subalgebra of
the SM lead to dangerous AdS vacua, as it happened in the non-SUSY SM. As in the latter, embedding the twist into hypercharge or
weak isospin lead only to  {\it Type S} vacua.  If we embed $Z_4$ into the Cartan subalgebra of $SU(3)$ the following 
particles get KK towers,

{$\bullet$ }  Two colours of quarks $\rightarrow \ 48$ d.o.f.

{$\bullet$ }  One colour  of squarks $\rightarrow \ 24$ d.o.f.

{$\bullet$ }   All sleptons  $\rightarrow \ 24$ d.o.f.

{$\bullet$ }  Gauge bosons, graviton  and Higgs  $\rightarrow \ 34$ d.o.f. 

The contribution of fermions and sfermions cancel neatly. 
 But altogether the number of bosons with a KK tower exceeds the number of fermions (since 8  gauginos 
do not get a KK tower, as well as Higgsinos). Therefore, as $a^2\rightarrow 0$ the potential necessarily develops an instability and the vacuum is of
{\it Type S}, see Fig. \ref{su3MSSM}.  Thus the MSSM, due to the presence of the sparticles,  survives the test that the SM failed.

The  dimension 4 action of the R-parity conserving MSSM has two family-independent global symmetries, one being $B-L$, like in the SM.
The other is an R-symmetry $U(1)_s$ with charges shown in Table \ref{cargas}. Let us consider them in turn.

\subsection{Embedding into  a discrete subgroup of $U(1)_{ B-L}$}

As we said, an embedding of $Z_N$ into  a discrete subgroup of $U(1)_{B-L}$  will always be possible if the MSSM is extended to
include a $U(1)_{B-L}$ gauge symmetry. 
This  symmetry is automatically present as a global symmetry  in the R-parity conserving MSSM, and indeed the $B-L$ symmetry is gauged in 
a multitude of string semi-realistic models (see e.g. \cite{bert}).  A discrete subgroup of $U(1)_{B-L}$ may also be gauged without the full $U(1)_{B-L}$ gauge
symmetry surviving at low energies. For our purposes there is no practical difference although of course, if the symmetry $U(1)_{B-L}$ is
fully gauged in nature our conclusions below would be completely  generic.

\begin{figure}
     \begin{center}
     	\subfigure[]{	
        \includegraphics[scale=0.38]{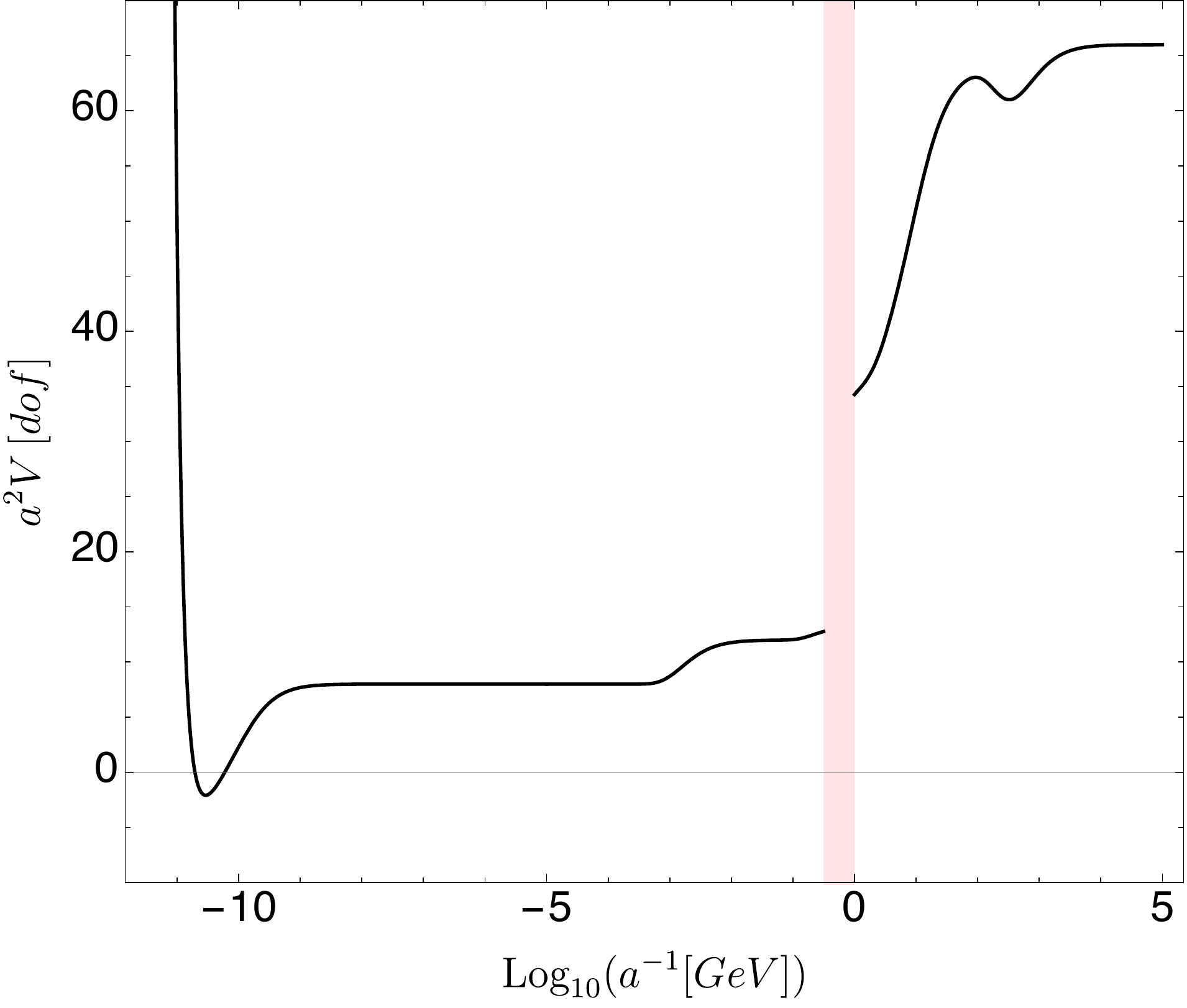}
     \label{SMAllperiodicleft}
        }
        \subfigure[]{
        \includegraphics[scale=0.38]{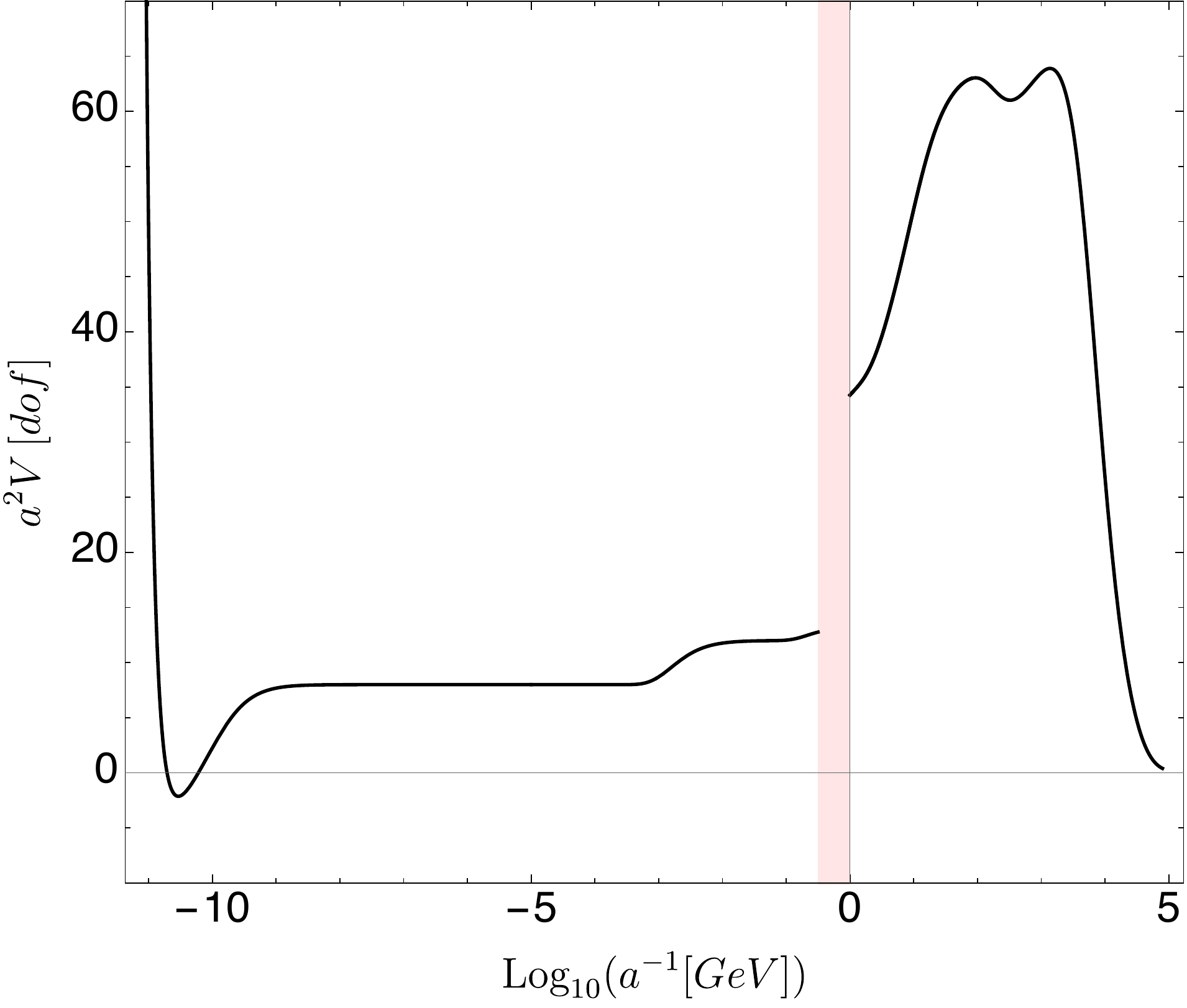}
     \label{SMAllperiodicright}
        }
      \caption{\footnotesize \textbf{(a)}  Effective potential of the MSSM compactified in $T^{2}/Z_{4}$ embedded into $B-L$ and the Higgs fixed at its minimum. The only possible 
      AdS minimum which may arise  is the one associated with neutrino bounds.  \textbf{(b)} Effective potential of the MSSM compactified in $T^{2}/Z_{4}$  and embedded into $U(1)_s$. The Higgs is fixed at its minimum. The only possible minimum is the one associated with neutrino bounds. Recall that 2D minima only appear when $V(a)$ crosses zero. The neutrino minimum may be avoided (in both cases) if the lightest Dirac neutrino is light enough.}
     \label{SMAllperiodic}
     \end{center}
\end{figure}

\begin{figure}
	\begin{center}
		\includegraphics[scale=0.40]{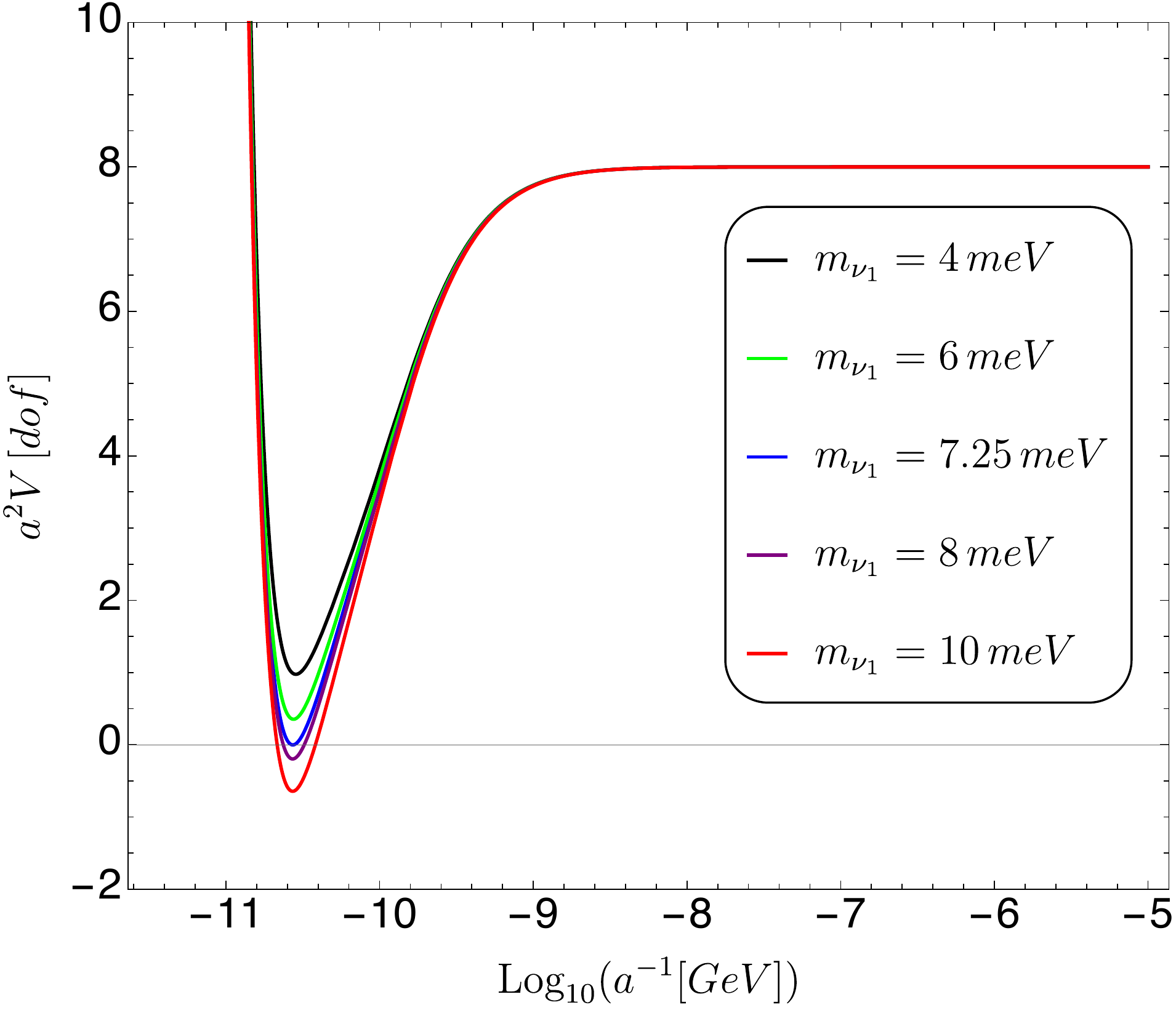}
		\caption{\footnotesize Zoom in of Fig. \ref{SMAllperiodic} around the neutrino scale. We always assume Dirac neutrinos, since Majorana neutrinos are ruled out. Using experimental constraints on the neutrino masses, we choose as free parameter the mass of the lightest neutrino. If its mass is larger than $7.25$ meV, an AdS minima develops. }
		\label{Zneutrino}
	\end{center}
\end{figure}

 The quantum numbers of each MSSM particle with respect to 
$(B-L)$ are shown in Table \ref{cargas}. Since the $q$ values shown in the table are all odd, all quarks and leptons get a KK tower, while sfermions do not. 
Gauginos and Higgsinos do not get KK towers since they do not have $B-L$ charge,  whereas Higgs scalars do have a KK tower.  The structure of the Casimir potential is, up to 
numerical factors with no important impact,  practically identical to 
the potential that we depicted for the case of the circle with Wilson lines fixed.   We show $a^2V(a)$ for this $Z_4$  case in 
Fig. \ref{SMAllperiodicleft}. We find one dS and (possibly) one AdS minima at the neutrino scale, depending on the mass of the lightest Dirac neutrino. Since there are more fermionic than bosonic degrees of freedom the potential keeps growing. This ensures the stability of the neutrino vacuum, eliminating the runaway solution that appears for the circle or the torus, where Wilson lines are present.  In Fig.  \ref{Zneutrino} we plot the potential around the neutrino scale as a function of the lightest Dirac neutrino, assuming Normal Hierarchy. We find that there is no minimum if its mass is lower than $7.25$ meV. For Inverse Hierarchy the bound is $m_{\nu_3}\leq 2.05$ meV. 
Thus we recover the possibility of an AdS vacuum
which is avoided for appropriate neutrino masses, and get back the predictions for the cosmological constant and the EW hierarchy.
We conclude that in the MSSM extended by $U(1)_{B-L}$ (or a discrete subgroup) one finds the attractive list of four predictions listed in the introduction. But this time without runaway directions that could spoil them.

Let us also note that in the SUSY case we do not have the freedom to set the torus boundary conditions for the leptons AP
and P for the quarks. The reason is that gauginos (and the gravitino) couple to both quarks and leptons so that one cannot chose
different torus boundary conditions for them. Thus the problematic AdS vacua present in 3D for the SM and shown in Fig. \ref{segment_periodic_SMright} is  not present in the SUSY case.
In this section 4 we are always taking periodic boundary conditions for all fermions. If all fermion boundary conditions were AP we would get uninteresting {\it Type S} vacua.

\subsection{Embedding into the $U(1)_s$ R-symmetry}

We finally consider the embedding of $Z_4$  into an order eight discrete subgroup of the $U(1)_s$ R-symmetry in Table \ref{cargas}.  R-symmetries of this type are also 
abundant in string theory \cite{discretestring}. In some cases they are gauged in the $\sigma$-model sense, and their anomalies are cancelled
by a Green-Schwarz mechanism, see e.g.\cite{kahleranomalies, discreteanomalies}.  
They are eventually broken upon SUSY breaking by dimension $<4$ operators.
Note that the table shows the charges of fermions, and that of scalars 
is obtained by adding (+1) and hence is always even.  It is easy to check that the complete spectrum of the MSSM get KK towers. 
In addition to the SM particles there will be contributions  from KK towers of
squarks, sleptons, gauginos, Higgsinos  and the gravitino.  We are thus including 48  (complex) scalars from the sfermions, 4 new Higgs (real)scalars, 4  Higgsinos, 12 gauginos and
one gravitino. Altogether an addition of 60 net bosonic degrees of freedom which just balances the mostly fermion dominated SM.
We show in Fig. \ref{SMAllperiodicright} the function $a^2V(a)$ for this case, in which we have chosen a specific SUSY mass spectrum, 
provided in Appendix \ref{app:MSSMspectrum}. The qualitative structure does not depend on the details of the spectrum. For values $1/a \gtrsim M_W$, the SUSY thresholds open up,
 and there is a cancellation between fermions and bosons so that the potential goes to zero as $a\rightarrow 0$.
Indeed we find that the leading contribution to the Casimir potential in this limit is given by
\beq
V\ \rightarrow \ \frac {1}{(2\pi a)^ 2} \frac {\mathcal G}{3} \left(\sum_P
(-1)^{2s_p+1}\ n_P\right) \ .
\label{pot4a0}
\eeq
which cancels if all MSSM particles have KK towers. Here ${\mathcal G}$ is Catalan's constant.  
One has to be slightly more careful, though.
Since the leading contribution as $a\rightarrow 0$  cancels, we have to examine the sub-leading terms to
decide whether a runaway behavior exists or not. In the case of the segment one would have for a SUSY spectrum from Eq. \eqref{limitesegment} a behaviour for small $R$
\beq
V(R)_{R\rightarrow 0}\ \rightarrow \ \frac {1}{48\pi R^4} \sum_p(-1)^{2s_{p}}n_pm_p^2\ \equiv \ \frac {1}{48\pi R^4} \delta_{SS} \ +\ {\cal O}(1/R^2)
\eeq
where $\delta_{SS}$ is the {\it supertrace} over all masses. In the torus and $T^2/Z_4$ cases something analogous but slightly more complicated is obtained.
The sub-leading mass-dependent contribution from a bosonic KK tower turns out to be (see Appendix \ref{app:1looptorusorbifold})
\beq
V^{(2)}\left[a,m,0\right]=  \left(am\right)^{2}\left\{ \pi\log\left(2\pi ma\right)-\frac{\pi^{2}}{6}-\frac{5\pi}{2}+\log\frac{\Gamma\left(\frac{1}{4}\right)}{2\pi^{3/4}}\right\} \label{eq:cuadraticperiodic}
\eeq
and the opposite for a fermion. In the  $(ma)\rightarrow 0$ limit we are considering the bosonic contribution is negative and the fermionic is
positive. Numerically the situation is rather similar to case of the segment. In particular, the $\log \left(2 \pi m a\right)$ gives even more importance to the most massive particles in the spectrum with respect to the case when only the supertrace appeared. When it comes to understanding the final sign of the potential it must be noted that, if the sign of the supertrace is dictated by the most massive particle in the spectrum (as it is usually the case), it is guaranteed that the sign of the potential will be actually given by the one of the supertrace.

There are then essentially two model-dependent possibilities, depending on the particular structure of the SUSY spectrum:
\begin{itemize}
\item
$\delta_{SS}\leq 0$. In this case there is again an instability at small $a$,  and we have a {\it Type S} vacuum.  The WGC constraints are evaded and there is no
inconsistency. 
\item
$\delta_{SS}>0$.  In this case there is no runaway direction and the possibility of stable neutrino minima is recovered,
as well as the corresponding predictions, we have a {\it Type P} vacuum. This time not only the neutrino masses are constrained but
also the SUSY masses.

\end{itemize}
Although both signs may lead to consistent theories, the $\delta_{SS}>0$ case is particularly attractive since the interesting constraints from the 
non-existence of neutrino AdS vacua are preserved.  The sign of the superstrace is model dependent. In particular it depends on the values of
the SUSY-breaking gaugino, squark, slepton and Higgssino/Higgs masses. One could think that in practically any MSSM constructed to date one has 
a  dominant positive contribution to $\delta_{SS}$, since there are many more massive SUSY bosons than SUSY fermions, due to family replication.
This is true, however, only if one restricts oneself to the effect of the SUSY partners of the observed SM particles. 
But here  it is also relevant the mass of the gravitino $m_{3/2}$. The minimum SUSY breaking sector should involve 
at least a goldstone chiral multiplet, with a goldstino and a s-goldstino.  The gravitino becomes massive combining with the goldstino and the s-goldstino typically also gets a
mass of the same order.
Thus one can also obtain $\delta_{SS}> 0$ by having e.g. $m_{sg}$ heavier than all the rest of the spectrum. The sign also obviously depends on the possible 
presence of further particles beyond the MSSM and gravitino/goldstino sectors.

Still, the condition $\delta_{SS}>0$ may be an important constraint on specific SUSY extensions of the SM. In particular, consider a  MSSM model in which the
gluino is the heaviest SUSY particle, larger than all other MSSM sparticles but also larger than $m_{3/2}$ and $m_{sg}$. If the gluino is heavy enough one will 
violate the $\delta_{SS}>0$ condition and, although possibly consistent with the AdS-phobia  condition, the radion potential would be unstable and the
predictions from the neutrino AdS vacua would be lost.  Note however that this can be avoided and the neutrino conditions would survive if e.g. the 
s-goldstino mass turns out to be heavier than the gluino.  Since the s-goldstino and gravitino sector depends strongly on each model we cannot 
get a firm prediction that the heaviest observable SUSY particle should be a boson, since the SUSY-breaking is in general precluded from the observable
sector. On the other hand  $\delta_{SS}> 0$ may be an interesting constraint to test in specific SUSY extensions of the SM.

\subsubsection{New AdS vacua for particular choices of SUSY spectra}

Although the $\delta_{SS}>0$ condition guarantees stability of the potential, it turns out that  there can be new  AdS minima 
at finite $1/a$ for particular choices of SUSY masses.  Let us show here a couple of examples. The general structure of these examples requires three ingredients. First, to have a boson as the heaviest particle in order to fulfill the $\delta_{SS}> 0$ condition and ensure local stability of the minimum. Second, to have an energy scale at which bosons dominate in order to be in the $V<0$ region and have a chance to form a minimum. Third, between this scale and the mass of the heaviest boson, we must have some fermionic degrees of freedom that can lift the potential to create the AdS minimum.  To make things simple, we will consider here all the supersymmetric particles to have masses around 1 TeV except  the boson that ensures $\delta_{SS}> 0$ and the fermion that gives a positive contribution to the potential.

\begin{figure}
     \begin{center}
     	\subfigure[]{	
        \includegraphics[scale=0.38]{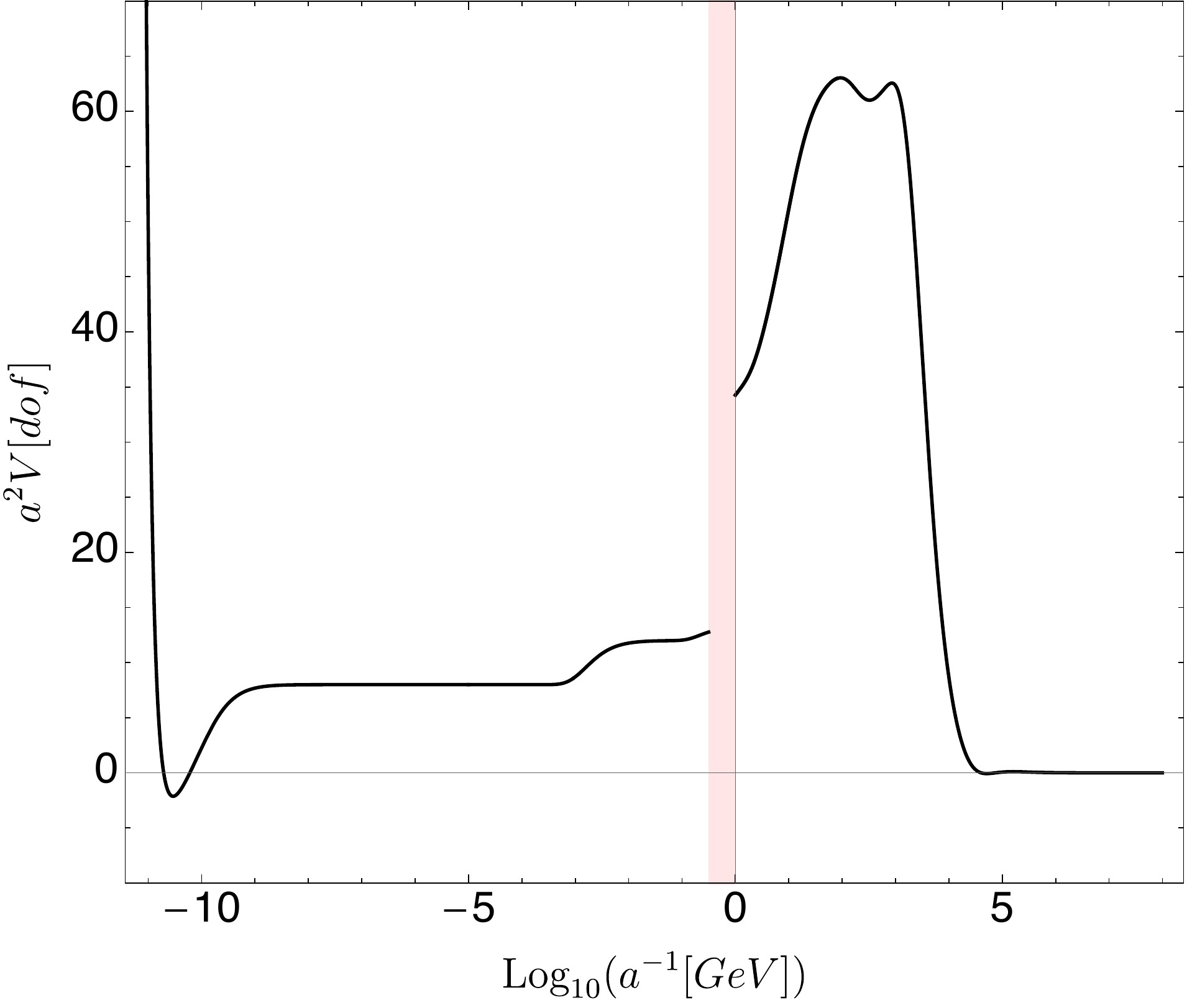}
   \label{SUSYconstraints1left}
        }
        \subfigure[]{
        \includegraphics[scale=0.38]{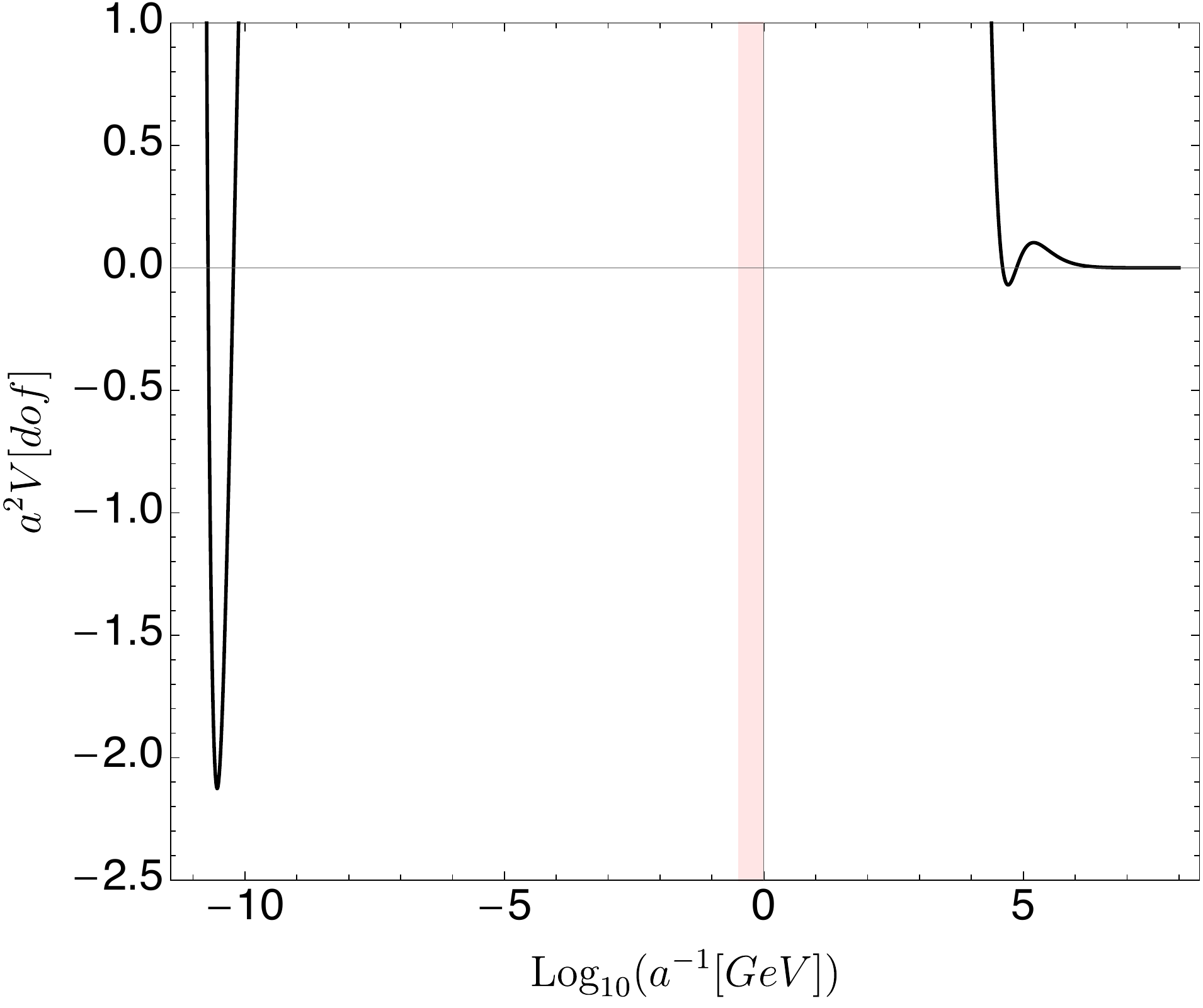}
  \label{SUSYconstraints1right}
        }
    \caption{\footnotesize Effective potential of the Standard Model compactified in $T^{2}/Z_{4}$  and embedded into $U(1)_s$. The masses of all superpartners of the SM particles are set to 1 TeV except for the gravitino (4 d.o.f.), $m_{3/2}=19 \ \mathrm{TeV}$ and the s-goldstino (2 d.o.f.), $m_{\text{sgolds}}=33 \ \mathrm{TeV}$. \textbf{(b)} displays a zoom in of \textbf{(a)} and it can be seen that a \textit{Type P} AdS minimum forms since it depends on the masses of the SUSY spectrum, hence giving some constraints from AdS-phobia.}
     \label{SUSYconstraints1}
     \end{center}
\end{figure}

\begin{figure}
     \begin{center}
     	\subfigure[]{	
        \includegraphics[scale=0.38]{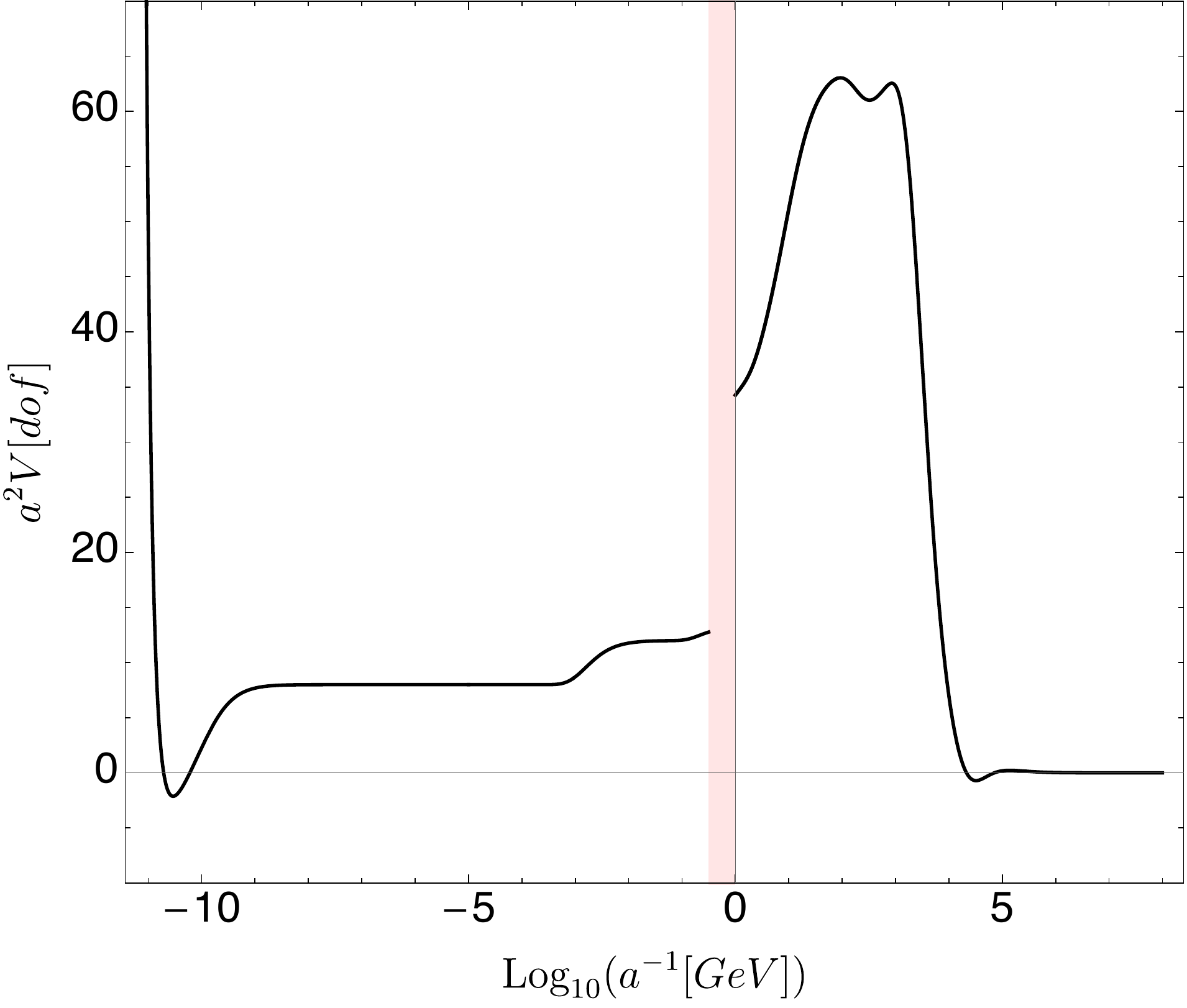} 
 \label{SUSYconstraints2left}
        }
        \subfigure[]{
        \includegraphics[scale=0.38]{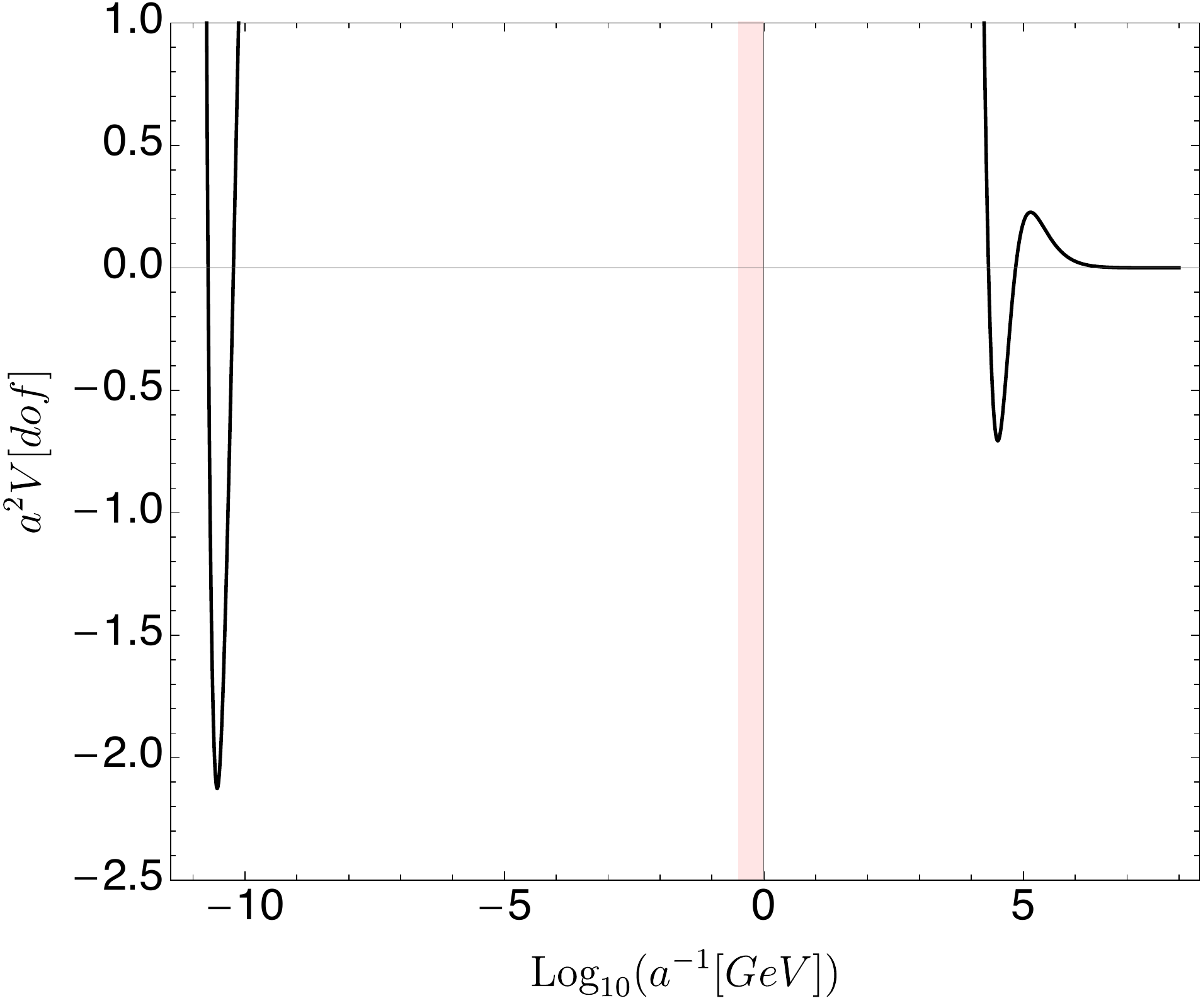}
 \label{SUSYconstraints2right}
        }
   \caption{\footnotesize  Effective potential of the Standard Model compactified in $T^{2}/Z_{4}$  and embedded into $U(1)_s$. The masses of all superpartners of the SM particles are set to 1 TeV except for the gluinos (16 d.o.f.), $m_{\text{g}}=18 \ \mathrm{TeV}$ and 1 squark (12 d.o.f.), $m_{\text{squark}}= 23 \ \mathrm{TeV}$. \textbf{(b)} displays a zoom in of \textbf{(a)} and it can be seen that a \textit{Type P} AdS minimum forms since it depends on the masses of the SUSY spectrum, hence giving some constraints from AdS-phobia.}
     \label{SUSYconstraints2}
     \end{center}
\end{figure}

The first example of these kind of vacua is shown in Fig. \ref{SUSYconstraints1} and we can constraint the relation between the mass of the gravitino and the mass of the s-goldstino (for fixed masses of all the other superpartners). In particular, an AdS minimum forms from $m_{3/2}= 19 \ \mathrm{TeV}$ and $m_{\text{sgolds}}=33 \ \mathrm{TeV}$ onwards, excluding these kind of spectra in SUSY models with $U(1)_S$ symmetries.
Another example of these kind of vacua can be seen in Fig. \ref{SUSYconstraints2}, in which we consider the most massive particle to be a squark of mass $m_{\text{squark}}=23 \ \mathrm{TeV}$ and the fermions that lift the potential to be the gluinos, with mass $m_{\text{g}}=18 \ \mathrm{TeV}$. From these masses on an AdS minimum forms and we can exclude these kind of spectra in the SUSY extensions of the SM.


More generally, given a SUSY version of the SM, computing the supertrace gives us information on the stability of the 
lower dimensional compactification and possible constraints  on the SUSY spectrum and other parameters, i.e. 
neutrino masses and the cosmological constant. Furthermore one has to check whether additional AdS vacua can form
depending on particular choices of SUSY masses.

\subsubsection{Charge-colour breaking AdS minima}

It is well known that the 4D  MSSM has, apart from the standard EW minima in which the Higgs fields get a vev,  a plethora 
of other possible minima in which squarks or sleptons are the ones who get a vev. \cite{frere,lleyda}. The space of these other minima is in general complicated and strongly
dependent on the SUSY-breaking parameters.
Many of these other minima are driven by the trilinear scalar couplings of the form e.g.  $A_th_t({\tilde t}_RH_u{\tilde t}_L)+h.c.$,  where 
$A_t$ is a soft parameter with dimension of mass, see \cite{lleyda} for a detailed analysis and references.  One can derive necessary conditions on
soft masses in order to avoid these minima to be lower than the standard Higgs minimum. A well known bound for minima derived from this trilinear
$stop$ coupling is 
\beq 
|A_t|^2 \ < \ 3h_t^2(m^2_{H_u}+m_{{\tilde t}_L}^2+m_{{\tilde t}_L}^2) \ .
\eeq
Weaker bounds may be derived by allowing minima lower than the Higgs one but imposing that the minimum is sufficiently stable at the 
cosmological level.  In fact most  of the examples of SUSY spectra discussed in the literature belong to this class in which the standard
Higgs MSSM vacuum is metastable.  These  charge/colour breaking minima   to  which the Higgs vacuum is unstable to decay are AdS 4D minima, 
and the deepest  of them will be stable. If we want to forbid AdS non-SUSY vacua altogether these 4D AdS minima should be absent from the start.

The logic of AdS-phobia then requires the Higgs vacua of the MSSM to be strictly stable and not just metastable. 
In fact full stability has been already advocated on different grounds, see e.g.\cite{hollik} and references therein for a recent 
discussion.  A general consequence of
imposing strict stability and a Higgs mass around 125 GeV is a quite heavy SUSY spectrum \cite{hollik}. On the other
hand this agrees well with the non-observation so far of any SUSY particle at LHC.

In going to 2D on  $T^2/Z_4$, a necessary condition in order not to get any of these AdS vacua inherited is to impose 
as a strict condition the  stability of the Higgs MSSM vacuum in 4D.  In addition, one expects that some of the stable vacua in 4D
may become unstable in 2D.  So the actual conditions on SUSY breaking mass parameters will be stronger than the parent 
4D conditions.  Thus one should impose that 1) the MSSM Higgs vacuum is stable in 4D against the decay into charge/colour breaking minima
and 2) it remains stable against decay in 4D.  Unfortunately these conditions can only be checked  in a case by case basis.

Summarizing, with the SM embedded into a SUSY completion the  AdS vacua which appeared in the non-SUSY SM become unstable 
and the theory is safe. The constraints for the neutrino masses, cosmological constant and EW hierarchy are recovered  from the
existence of a $Z_4$ vacua with action embedded into a discrete subgroup of $B-L$ in the R-parity preserving MSSM.
Further constraints on the SUSY spectrum appear from a $Z_4$ embedding into a discrete subgroup of the R-symmetry 
present in the R-parity preserving MSSM, if the supertrace $\delta_{SS}>0$. 
Note that the latter constraint  on the MSSM spectrum would only arise if the discrete subgroup of $U(1)_s$ is a gauge symmetry 
of the underlying theory.

We see that the EW scale is bounded to be close to its experimental value in order to avoid the generation of stable 2D AdS neutrino generated vacua.
On the other hand,  avoiding   4D and/or 2D AdS charge-colour breaking minima typically requires relatively  large 
SUSY masses in the multi-TeV region.  Note that this could explain the so called {\it little hierarchy problem}:  the SUSY spectrum needs to be
relatively heavy in order to avoid these dangerous minima. On the other hand the EW scale is kept smaller by the condition that the 
lightest neutrino is sufficiently light to avoid AdS neutrino vacua.  
We must emphasize, though, that from this discussion the scale of SUSY breaking needs not be 
in the multi-TeV region. It could be much larger, up to a scale e.g. of order $10^{10}$ GeV and the stability properties would still persist.

\subsection{Twisted sectors}

In all the previous sections  we have not discussed the possible existence of twisted sectors. The presence of those 
is expected if there are orbifold fixed points in the compactification.  Fixed points may be absent if the twist in the torus
is accompanied by some translation in the compact six dimensions of the original 4D compactification. 
If they are present, in general there  will be  2D particles from twisted sectors. Thus e.g. they explicitly arise in heterotic orbifold 
compactifications to 2D (see  e.g. \cite{anamaria, 2Danomalies}). In Type II orientifolds there may be branes (D-strings in the 2D case) localized 
on the fixed points, leading to additional massless particles in the 2D theory.  A full discussion of these twisted sectors would require a 
full knowledge of the underlying string compactification which leads to the
SM or the MSSM in the first place.  Fortunately for our purposes, to leading order we can ignore the effect of possible twisted sectors in our
Casimir energy computations.
This is because any twisted particle is localized in the singularities and hence does not have any KK tower.  As we have emphasized,
only states with KK towers contribute to the Casimir potential and hence twisted sectors do not modify to leading order the
structure of the radius potential.  In this respect their effect is similar to untwisted zero modes which also do not contribute to the
Casimir potential.  Notice however that in principle a radius independent constant contribution could arise from the twisted sector.
This could be interpreted as a contribution of the tension of the objects localized in the singularities. Like e.g. in orientifold compactifications
in which the tension of branes and orientifolds cancel, one may expect such local contribution to cancel at leading order. 
We are assuming that vacua exist in which  such constant contribution is either absent or small.
A small constant term would not alter significantly our discussion. At large $a$ the bulk 4D  $\Lambda_4$  clearly dominates (it is multiplied by $a^2$) whereas
at small $a$ the Casimir energy contributions grow rapidly like $1/a^2$. Therefore, a constant piece may postpone the growth of 
a  negative potential due to the photon and the graviton but it cannot avoid the appearance of an AdS minima.

The compactified models here discussed  are in general chiral in 2D. Thus e.g. the model constructed by twisting by a a $Z_4$ 
subgroup of $B-L$ have zero modes transforming like quarks  under the SM group, with different 2D chiralities and has $SU(2)_L$ 2D anomalies.
The model obtained by twisting by a $U(1)_s$ subgroup also has chiral zero modes transforming like all fermions of the MSSM. 
In 2D there are in general gauge and gravitational anomalies \cite{AlvarezGaume} and indeed these spectra by themselves have 2D anomalies. 
In particular, gauge anomalies are given by the quadratic Casimir eigenvalue $T_a$ of each chiral fermion, with sign depending on whether the
zero mode is left- or right-moving.  In principle one could use 2D anomaly cancellation conditions to try to figure out what
the quantum numbers of the 2D twisted sectors could be. Indeed, one may easily obtain anomaly-free 2D theories by adding 
appropriate representations, which would be 2D lepton-like objects in our case.
See e.g. \cite{anamaria} for specific 2D examples of anomaly cancellation.
Although, as discussed above, the twisted sectors play no role in the computation of the Casimir potential, it would be interesting to study
further the 2D anomaly constraints in specific 2D compactifications of the SM, MSSM or generalizations, i.e. compactifying 
a 10D theory directly to 2D.  We leave this for future work.

\section{Conclusions and outlook}
\label{sec:conclusions}

In this paper we have studied compactifications of the SM to 3D and 2D, looking for stable AdS vacua,
completing and generalizing previous work in \cite{ArkaniHamed:2007gg, IMV1}. In those works
it was shown how the Casimir energy of the lightest sector of the SM gives rise to a radius dependent 
potential which may have AdS minima.
Our motivation was the
conjecture in \cite{OV} that posits that no theory with stable, non-SUSY AdS vacua can be embedded into a
consistent theory of quantum gravity.  In \cite{IMV1} constraints on neutrino masses were obtained from
this condition applied to the SM compactified on the circle and on the torus. Here the assumption of background independence is 
crucial so that the constraints can be applied to the theory obtained upon compactification. This is also an important assumption in the present paper.
For the minimal SM one finds that 
neutrinos cannot be Majorana (as in fact already suggested in \cite{OV}) and upper  bounds on the lightest neutrino mass
$m_{\nu_1}\leq 4.1\times 10^{-3}$ eV(NH) or $m_{\nu_3}\leq 1\times 10^{-3}$ eV(IH) are obtained.
Furthermore, it was found that the 4D cosmological constant is bounded from below by the scale of  neutrino masses,
$m_{\nu_1}^4\lesssim \Lambda_4$. This is an attractive prediction, since it is the first argument implying a 
non-vanishing $\Lambda_4$ only on the basis of particle physics, with no cosmological input.  Finally, 
the upper bound on the neutrino mass implies an upper bound on the EW scale (for fixed Yukawa coupling),
Eqs. (\ref{jerarquia}), (\ref{hierarchy}).
This could give an explanation for the stability of the Higgs mass against quantum corrections, i.e.
the hierarchy problem. Larger values of the EW scale allow for the generation of  AdS vacua.

Although the  AdS minima which may arise are perturbatively stable,   non-perturbative instabilities  could arise 
towards decay into lower minima or runaway directions, if present  at smaller compact radius $R<1/m_e$, with $m_e$ the
electron mass \cite{HS}. Indeed,  as we decrease $R$,  new thresholds of leptons and quarks become relevant to the
Casimir energy and the potential becomes more and more complex. In particular, it was shown that 
the presence of photon Wilson line moduli may give rise to non-perturbative instabilities \cite{HS}. If such instabilities exist,
there would be no contradiction with AdS-phobia, although then we would lose the attractive constraints summarized above.

Searching for predictive vacua, 
in this paper we study SM compactifications in which  Wilson line moduli of the SM gauge group are 
projected out. This is known to happen in orbifold $T^2/Z_N$ compactifications and we consider in particular
the $Z_4$ case because of its simplicity and  because for $N>2$ the complex structure of the torus is also projected out and we are left
with a potential depending only on the area of the torus and the Higgs scalar. We classify the different vacua we find as {\it Type D}, which contain a stable AdS minimum which cannot be avoided by constraining free parameters of the theory, {\it Type S}, which have no stable AdS minimum and {\it Type P}, which have  AdS vacua or not depending on some free parameters of the theory.
Twisting only by $Z_N$ gives rise only to {\it Type S} vacua and no constraints. More interesting vacua are obtained if we embed the $Z_4$ symmetry into
$U(1)$ gauge (or gaugable) degrees of freedom of the theory.  Embedding $Z_4$ into the SM gauge group we obtain necessarily some stable AdS vacua for the SM.
Thus, remarkably, the SM as it stands would not be embeddable into a consistent theory of quantum gravity.  This is true for any
value of neutrino masses. 
Note that  a further problem for the minimal SM is the possible   existence of 
a second 4D  high energy Higgs vaccum at scales  above $10^{10}$ GeV. This minimum would be AdS
and, if stable, would again be in contradiction with the OV conjecture.

Interestingly, if the SM is embedded into a SUSY version like the MSSM, those AdS vacua become automatically unstable and the theory is consistent with 
WGC constraints.  Furthermore, the second high energy Higgs vacuum also disappears if the SUSY breaking scale is not above $\sim 10^{12}$ GeV.
We study further possible {\it Type P} vacua of the MSSM compactified in $T^2/Z_4$ which could lead to constraints on particle physics.
In particular, an interesting vacuum is obtained if we embed $Z_4$ into a discrete subgroup of the $U(1)_{B-L}$ symmetry, which is a global symmetry of the 
R-parity conserving MSSM and may be gauged at higher energies. The resulting vacuum has a potential with just a possible  AdS minimum around the neutrino region.
This is the same AdS minimum  found in \cite{ArkaniHamed:2007gg,IMV1}, and may be avoided if one of the (Dirac) neutrinos is sufficiently light.
Thus we again have the constraints on the cosmological constant and the gauge hierarchy described in \cite{IMV1,IMV2}.  This suggests that 
the MSSM should be extended to include a $U(1)_{B-L}$ gauged symmetry.

It is interesting to see whether there are other $Z_4$ compactifications leading to further phenomenological restrictions e.g., on the
values of the SUSY masses.
The other family-independent global symmetry of the R-parity preserving MSSM is the R-symmetry $U(1)_s$ of Table \ref{cargas}. Such type of R-symmetries are gauged in the
$\sigma$-model sense in $N=1$ supergravity theories obtained in string compactifications. If we embed $Z_4$ into a discrete subgroup of this 
$U(1)_s$ we obtain a 2D model in which all MSSM particles have a KK tower. In this case, depending (essentialy) on the mass$^2$ supertrace, 
new conditions on the SUSY spectrum appear. However in this latter case the constraints would apply only  if the underlying theory contains 
such discrete R-symmetry.

Additional constraints come from avoiding the presence of charge and/or colour breaking  AdS minima 
both in 4D and 2D.  This typically requires a relatively heavy SUSY spectrum in the multi-TeV region \cite{hollik}.
These arguments could provide a possible explanation for the {\it little hierarchy problem}, i.e. the fact that,
if low energy SUSY is correct,  the 
SUSY particles seem to be relatively heavy compared to the EW scale. From the present point of view the EW scale
is small due to the constraint in Eqs.(\ref{jerarquia}), (\ref{hierarchy}) whereas the SUSY spectrum is forced to be relatively large to avoid
charge/colour breaking AdS minima.

An important question is whether the scale of SUSY-breaking  $M_{SS}$ and the masses of SUSY particles  are close to the EW scale and LHC energies or not.
It would be extremely interesting if the AdS-phobia condition as applied to the MSSM could give us a hint on what the scale of SUSY breaking is.
In particular it is conceivable e.g. that avoiding the appearance of AdS vacua when compactifying  the MSSM  down to 2D could
require a low-energy SUSY spectrum around a few TeV, low-energy SUSY.  Or else that avoiding such vacua could require a very massive SUSY spectrum.
Note in this respect that, as we said,  SUSY cannot be arbitrarily high, since if $M_{SS}> 10^{10}$ GeV the second lower SM Higgs minimum may develop,
which would be in AdS.  
To test whether there is a preference for low-energy SUSY or not coming from AdS-phobia we have to improve the present analysis. 
In particular we should consider the renormalization group improved couplings and masses. When going to very high energies (or rather very small $a$) 
large logs will appear which cannot be ignored at the quantitative level. The soft masses run and e.g the sign of $\delta_{SS}$ may change 
as masses and couplings run. We leave this important analysis for future work.

We have used the 2D vacua in this paper as auxiliary tools in order to derive constraints on the parent SM or MSSM 4D model. However 
some of these 2D vacua could be cosmologically interesting in the following sense. Consider the 2D vacua obtained from embedding into $U(1)_s$.
It contains zero modes for all the particles in the MSSM except for the Higgs bosons. 
Once the AdS neutrino (and/or SUSY) bounds are respected, the potential is monotonously decreasing into large 2D volume, the KK towers become massless
and we recover the 4D MSSM (including the Higgs bosons) as $a\rightarrow \infty$. So one could speculate that the universe started two-dimensional and became 4D well before 
experimentally constrained cosmological events took place. In such a model the 4D cosmological constant would have been 2D volume dependent.
The situation would be consistent with the conjecture  in \cite{vafafederico} that no stable   dS vacuum  should exist and the universe should have a runaway behavior. It would be 
interesting to explore whether a sensible cosmology could be constructed in this scheme.

Since the casuistic above could confuse the reader,  let us conclude with a brief list our findings:
\begin{itemize}
	\item The SM as it stands  necessarily has stable AdS vacua in 2D and hence would be in the swampland.
	\item In the MSSM those AdS vacua become unstable and lead to no incompatibility with quantum gravity.
	\item If the MSSM is extended by a $U(1)_{B-L}$ gauge group (or a discrete subgroup),  the theory has 2D AdS vacua
	which can be avoided if neutrinos are Dirac and the lightest  is sufficiently light. The four predictions listed in the introduction are
	recovered.
	\item If in addition the MSSM 4D vacua has a gauged discrete R-symmetry, subgroup of the global $U(1)_s$ R-symmetry of the 
	R-parity preserving MSSM, further constraints on the SUSY spectra, depending on the supertrace, are obtained
	\item The hierarchy problem is solved by imposing absence of AdS neutrino vacua. But SUSY is needed to 
	avoid additional AdS minima which would otherwise be present. Thus the SUSY spectrum could be substantially above the 
	EW scale, but also possibly in the multi-TeV region.
\end{itemize}

Finally, it would be important to improve our understanding of the stability of this kind of SM compactifications to lower
dimensions.  In particular in the 2D vacua
the radius $a$ does not propagate and the Higgs is the only propagating degree of freedom (along with the sfermions if present).
It would be interesting to study how the tunneling towards lower minima or runaway directions happens in this class of theories.
More generally the results depend on the validity of the assumption of the Ooguri-Vafa conjecture of AdS-phobia and it would be
important to bring additional evidence in favour or against it. In the meantime we think it is well motivated to 
study what the consequences of its validity would be, and they look, indeed, quite intriguing.

\vspace{1.0cm}

\newpage

\centerline{\bf \large Acknowledgments}

\bigskip

\noindent We thank A. Font, F. Marchesano, V. Martin-Lozano, A. Uranga,   I. Valenzuela, C. Vafa and especially M. Montero  for useful discussions 
and suggestions. 
This work has been supported by the ERC Advanced Grant SPLE under contract ERC-2012-ADG-20120216-320421, by the grant FPA2012-32828 from the MINECO,  and the grant SEV-2012-0249 of the ``Centro de Excelencia Severo Ochoa" Programme.   A.H. is supported by the Spanish FPU Grant No. FPU15/05012 and E.G. by the Spanish FPU Grant No. FPU16/03985.

\newpage

\appendix

\section{One-Loop Effective Potential in a $S^1$ compactification of the Standard Model.}
\label{app:1loopcircle}

We denote the quantum effective action by $\Gamma$ and the classical,
background fields by an overline. We are interested only in the ground
state of the theory. For this reason we can set the classical, background
fields of the fermions to zero from the start. We perform the computation
in the background-field gauge, because we will use the background-field
method to compute the effective action and this choice enables us
to maintain gauge invariance in the background, gauge-boson fields.
If we parameterize the Higgs doublet as $\frac{1}{\sqrt{2}}\eta+\frac{1}{\sqrt{2}}\left(\begin{array}{c}
1\\
\text{v}
\end{array}\right)$, then the gauge
fixing term is given by:
\begin{align}
{\cal {\cal L}}_{\text{Gauge Fixing}} & =-\frac{1}{2\xi}[\partial_{\mu}W_{i}^{\mu}+g\varepsilon_{ijk}\overline{W}_{\mu j}W^{\mu k}+i\frac{\xi}{2}g(\eta^{\dagger}\frac{\sigma_{i}}{2}\text{v}-\text{v}^{\dagger}\frac{\sigma_{i}}{2}\eta)]^{2}\nonumber\\
&\quad-\frac{1}{2\xi}[\partial_{\mu}B^{\mu}+\frac{i\xi g'}{4}(\eta^{\dagger}Y\text{v}-\text{v}^{\dagger}Y\eta)]^{2}-\frac{1}{2\alpha}[\partial_{\mu}G_{a}^{\mu}+gf_{abc}\overline{G}_{\mu b}G^{\mu c}]\label{eq:gauge_fixing}.
\end{align}
We work in the unitary gauge, obtained by taking the limit $\xi,\alpha\rightarrow\infty$.
Using the available background gauge invariance one can gauge away
all components except the one along the compact direction of those
bosons in the Cartan Subalgebra of $SU(3)_{C}\times SU(2)_{L}\times U(1)_{Y}$.
We denote these four Wilson lines as $\overline{Z},\overline{A},\overline{G}_{1},\overline{G}_{2}$,
corresponding to the vevs of the Z boson, the photon and the gluons associated with the two diagonal Gell-Mann matrices. We define
the Wilson lines so that they absorb the gauge coupling constants
in the covariant derivative.
Using the background field method, we must compute the following path
integral, neglecting terms associated with diagrams which are not
connected and 1PI. At one-loop order we need only study the terms quadratic
in the quantum fields, since linear terms would never give 1PI diagrams
and zero order terms correspond to the classical (tree level) action.
Denoting by $\Psi$ to all fermions and by $\eta$ to all the necessary
ghosts, and denoting in general as $\phi$ to all fields in the theory,
the measure would then be $D\phi=Dg_{\mu\nu}DA_{\mu}DZ_{\mu}DW_{\mu}^{+}DW_{\mu}^{-}DH\,D\overline{\Psi}D\Psi D\overline{\eta}D\eta$
and the path integral to compute
\begin{align}
e^{i\Gamma[\overline{R},\overline{H},\overline{Z},\overline{A},\overline{G}_{1},\overline{G}_{2}]} & =\int D\phi\;e^{iS[\phi+\overline{\phi}]}.\\
& \quad\text{1PI, Connected}\nonumber
\end{align}
For completeness, we will write in some detail the SM Lagrangian.
We will omit from the Lagrangian the part of the gauge-fixing term
\ref{eq:gauge_fixing} that depends on the background fields (of course,
it must included in the calculations of the effective action). We
do not write in any detail the ghost Lagrangian ${\cal {\cal L}}_{\text{FP}}^{SM}$,
since it was shown in \cite{Appelquist} that they do not contribute
to the effective action at one-loop order. The computation of the
effective potential of the Einstein-Hilbert Lagrangian compactified
in $S^{1}$, to one-loop order, was performed in \cite{Appelquist},
so we also omit the details here. We will also omit the details of
the counterterm Lagrangian and the renormalization procedure. To regularize
we will use either Dimensional Regularization with extra flat dimensions
or Zeta Function Regularization techniques. Note that $\mu,\nu,n,m$
space-time indices run from $0,1,2,3$; $i,j$ space-time indices
from $0,1,2$ ; $i,j$ colour indices run from $1,2,3$ and $a,b$
colour indices run from $1,2...8$. We will replace all space-time
covariant derivatives by partial derivatives from the start, since
we will be interested only in the one-loop corrections to the tree level
potential. Finally, $f$ denotes sum over all fermions and $A$ sum
over the fermions families: $e_{A}=(e,\mu,\tau)$, $\left(\begin{array}{c}
p_{A}\\
n_{A}
\end{array}\right)=$$\left(\begin{array}{ccc}
u & c & t\\
d & s & b
\end{array}\right)$.
\[
{\cal L}={\cal {\cal L}}_{\text{asymptotic}}^{SM}+{\cal {\cal L}}_{\text{basic interaction}}^{SM}+{\cal {\cal L}}_{\text{FP}}^{SM}+{\cal {\cal L}}_{\text{counterterms}}^{SM}+{\cal {\cal L}}_{\text{Einstein-Hilbert}}+\Lambda_{4}
\]
\[
{\cal {\cal L}}_{\text{asymptotic}}^{SM}={\cal {\cal L}}_{\text{Fermions}}+{\cal {\cal L}}_{\text{YM}}^{(2)}+{\cal {\cal L}}_{\text{SBS}}^{(2)}
\]
\[
{\cal {\cal L}}_{\text{Fermions}}=\sum_{f}\bar{\Psi}_{f}(i\tilde{\gamma}^{\mu}\partial_{\mu}-m_{f})\Psi_{f}\qquad\qquad\tilde{\gamma}^{\mu}=e_{n}^{\mu}\gamma^{n}\qquad g_{\mu\nu}=\eta_{nm}e_{\mu}^{n}e_{\nu}^{m}
\]
\begin{align*}
{\cal {\cal L}}_{\text{YM}}^{(2)} & =W_{\mu}^{-}[\eta^{\mu\nu}\partial^{2}-\partial^{\mu}\partial^{\nu}+\eta^{\mu\nu}M_{W}^{2}]W_{\nu}^{+}+\frac{1}{2}Z_{\mu}[\eta^{\mu\nu}\partial^{2}-\partial^{\mu}\partial^{\nu}+\eta^{\mu\nu}M_{Z}^{2}]Z_{\nu}\\
& \quad+\frac{1}{2}A_{\mu}[\eta^{\mu\nu}\partial^{2}-\partial^{\mu}\partial^{\nu}]A_{\nu}+\frac{1}{2}G_{\mu}^{a}[\eta^{\mu\nu}\partial^{2}-\partial^{\mu}\partial^{\nu}]\delta_{ab}G_{\nu}^{b}
\end{align*}
\[
{\cal {\cal L}}_{\text{SBS}}^{(2)}=-\frac{1}{2}H[\partial^{2}+m_{H}^{2}]H
\]
\[
{\cal {\cal L}}_{\text{basic interaction}}^{SM}={\cal {\cal L}}_{\text{NC}}+{\cal {\cal L}}_{\text{CC}}+{\cal {\cal L}}_{\text{YM}}^{3}+{\cal {\cal L}}_{\text{YM}}^{4}+{\cal {\cal L}}_{\text{SBS}}^{(3+4)}+{\cal {\cal L}}_{\text{YW}}
\]
\[
{\cal {\cal L}}_{\text{NC}}=\sum_{f}\bar{\Psi}_{f}\tilde{\gamma}^{\mu}(eQ_{f}A_{\mu}+\frac{g}{c_{W}}(g_{L}^{f}P_{L}+g_{R}^{f}P_{R})Z_{\mu}+g_{s}G_{\mu}^{a}T_{f}^{a})\Psi_{f};\quad T_{f}^{a}=\left\{ \begin{array}{c}
\,0\quad\text{colour singlet}\\
\frac{\lambda^{a}}{2}\quad\text{colour triple}
\end{array}\right\} 
\]
\[
{\cal {\cal L}}_{\text{CC}}=\frac{g\tilde{\gamma}^{\mu}}{\sqrt{2}}\sum_{A}\overline{\nu}_{A}W_{\mu}^{+}P_{L}e_{A}+\overline{p}_{A}W_{\mu}^{+}P_{L}n_{A}+h.c
\]
\begin{align*}
{\cal {\cal L}}_{\text{YM}}^{3} & =igc_{w}[(\partial_{\mu}W_{\nu}^{-}-\partial_{\nu}W_{\mu}^{-})W^{+\mu}Z^{\nu}-h.c]-ie[(\partial_{\mu}W_{\nu}^{-}-\partial_{\nu}W_{\mu}^{-})W^{+\mu}A^{\nu}-h.c]\\
& \quad+igc_{w}(\partial_{\mu}Z_{\nu}-\partial_{\nu}Z_{\mu})W^{+\mu}W^{-\nu}-ie(\partial_{\mu}A_{\nu}-\partial_{\nu}A_{\mu})W^{+\mu}W^{-\nu}\\
& \quad-\frac{1}{2}g_{s}f_{abc}(\partial_{\mu}G_{\nu}^{a}-\partial_{\nu}G_{\mu}^{a})G^{\mu b}G^{\nu c}
\end{align*}
\begin{align*}
{\cal {\cal L}}_{\text{YM}}^{4} & =-g^{2}c_{w}^{2}[W_{\mu}^{-}W^{+\mu}Z_{\nu}Z^{\nu}-W_{\mu}^{-}W_{\nu}^{+}Z^{\mu}Z^{\nu}]-e^{2}[W_{\mu}^{-}W^{+\mu}A_{\nu}A^{\nu}-W_{\mu}^{-}W_{\nu}^{+}A^{\mu}A^{\nu}]\\
& \quad+egc_{w}[2W_{\mu}^{-}W^{+\mu}Z_{\nu}A^{\nu}-W_{\mu}^{-}W_{\nu}^{+}(Z^{\mu}A^{\nu}+A{}^{\mu}Z^{\nu})]\\
& \quad+g^{2}(W_{\mu}^{-}W^{+\mu}W_{\nu}^{-}W^{+\nu}-W_{\mu}^{+}W^{+\mu}W_{\nu}^{-}W^{-\mu})+\frac{1}{4}g_{s}^{2}f_{abc}f_{ade}G_{\mu}^{b}G_{\nu}^{c}G^{\mu d}G^{\nu e}
\end{align*}
\[
{\cal {\cal L}}_{\text{SBS}}^{(3+4)}=\frac{g^{2}}{4}[2\text{v}H+H^{2}][W_{\mu}^{-}W^{+\mu}+\frac{1}{2c^{2}_{w}}Z_{\mu}Z^{\mu}]-\lambda\text{v}H^{3}-\frac{\lambda H^{4}}{4}
\]
\[
{\cal {\cal L}}_{\text{YW}}=-\sum_{f}m_{f}\frac{H}{\text{v}}\bar{\Psi}_{f}\Psi_{f}.
\]

\[
g_{L,R}^{f}=T_{3\;L,R}^{f}-s_{w}^{2}Q_{f}
\]
\begin{table}
	\begin{centering}
		\begin{tabular}{|c|c|c|c|}
			\hline 
			& $Q_{f}$ & $g_{L}^{f}$ & $g_{R}^{f}$\tabularnewline
			\hline 
			\hline 
			$\nu_{e}$,$\nu_{\mu}$,$\nu_{\tau}$ & $0$ & $\frac{1}{2}$ & $0$\tabularnewline
			\hline 
			$e$,$\mu$,$\tau$ & $-1$ & $-\frac{1}{2}+s_{w}^{2}$ & $s_{w}^{2}$\tabularnewline
			\hline 
			$u$,$c$,$t$ & $\frac{2}{3}$ & $\frac{1}{2}-\frac{2}{3}s_{w}^{2}$ & $-\frac{2}{3}s_{w}^{2}$\tabularnewline
			\hline 
			$d$,$s$,\textbf{$b$} & $-\frac{1}{3}$ & $-\frac{1}{2}+\frac{1}{3}s_{w}^{2}$ & $\frac{1}{3}s_{w}^{2}$\tabularnewline
			\hline 
		\end{tabular}\caption{$SU(2)_{L}\times U(1)_{Y}$ quantum numbers of the particles in the
			Standard Model.}
		\par\end{centering}
\end{table}

Taking into account the boundary conditions, each field is expanded in a Fourier series, as explained in the text. Since we are staying at cuadratic order, the integral over each field will be gaussian.
Each gaussian integral reduces to computing the determinant of a certain operator. Using $ \det A\text{=}e^{\log\text{Tr}A}, $
we transform the determinant into a trace. After computing the traces for the different particles 
one finds that the contribution of each particle to the
Casimir potential can always be written as:
\begin{equation}
V_{p}=(-1)^{2s_{p}+1}n_{p}\frac{-i}{2}\sum_{n=-\infty}^{\infty}\int\frac{d^{3}p}{(2\pi)^{3}}\log\left(-p^{2}\varphi^{2}+\left(m_{p}\frac{H}{\text{v}}\right)^{2}+\left(n+\theta\right)^{2}\varphi^{2}\right).
\end{equation}
\section{One-Loop Effective Potential in $T^{2}/Z_{4}$.}
\label{app:1looptorusorbifold}

The first steps of the computation are essentially the same as in the previous Appendix. Besides, some details were given in the text so we do not repeat them here. 
In the text we also mentioned that, to gain a better understanding of the Casimir potential, it is interesting to study 
the ultraviolet behaviour. In this Appendix we will use the expressions from the text to extract this UV behaviour. To take this $am\rightarrow 0$ limit, it will be useful to rewrite the general formula for the Casimir potential in the torus using the following formulas from \cite{Lewin}
\beq
\text{Li}_{s}(re^{i\phi})=\text{Li}_{s}(r,\phi)+i\left[\text{Ti}_{s}\left(\frac{r\sin\phi}{1-r\cos\phi}\right)-\text{Ti}_{s}\left(\frac{r\sin\phi}{1-r\cos\phi},\tan\phi\right)\right],\qquad s=2,3
\eeq
where 
\beq
\text{Li}_{2}(r,\phi)=-\frac{1}{2}\int_{0}^{r}\frac{\log \left(1-2x\cos\phi+x^{2}\right)}{x}dx,
\eeq
\beq
\text{Li}_{3}(r,\phi)=\int_{0}^{r}\frac{\text{Li}_{s}(x,\phi)}{x}dx,
\eeq
and $\text{Ti}_{s}$ are generalized inverse tangent integrals which
verify: $\text{Ti}_{s}\left(-x\right)=-\text{Ti}_{s}\left(x\right)$,
$\text{Ti}_{s}\left(-x,-y\right)=-\text{Ti}_{s}\left(x,y\right)$.
Thus, $\text{Li}_{s}(re^{i\phi})+\text{Li}_{s}(re^{-i\phi})=2\text{Li}_{s}(r,\phi).$
Introducing this last equation in  \eqref{toruscasimir} we find:
\begin{align}
V_{\mathcal{C}}\left[a,t,m_{p},\theta_{1},\theta_{2}\right] & =\frac{1}{\left(2\pi a\right)^{2}}\left[2a^{2}t_{2}m^{2}\sum_{p=1}^{\infty}\frac{\cos2\pi p\theta_{1}}{p^{2}}K_{2}\left(\frac{2\pi pam}{\sqrt{t_{2}}}\right)\right.\nonumber\\
& \left.+\frac{1}{2\pi t_{2}}\sum_{n=-\infty}^{\infty}\left\{ -\log \,r_{n}\text{Li}_{2}(r_{n},\phi_{n})+\text{Li}_{3}(r_{n},\phi_{n})\right\} \right],\label{eq:simplifytorus}
\end{align}
\begin{equation}
e^{\sigma_{+}}=e^{-2\pi\sqrt{\left(n+\theta_{1}\right)^{2}t_{2}^{2}+m^{2}a^{2}t_{2}}}e^{i2\pi\left[-\left(n+\theta_{1}\right)t_{1}+\theta_{2}\right]}=re^{i\phi}.\label{eq:apoyosimplytorus}
\end{equation}
Secondly, we set $t_{1}=0$ and $t_{2}=1$ in Eq. \eqref{eq:simplifytorus}
and Eq. \eqref{eq:apoyosimplytorus}, $\theta_{1}=\theta_{2}\equiv\theta$,
and take the massless limit:
\begin{align}
V_{\mathcal{C}}\left[a,\theta\right] & =\frac{1}{\left(2\pi a\right)^{2}}\left[\frac{1}{\pi^{2}}\sum_{p=1}^{\infty}\frac{\cos2\pi p\theta}{p^{4}}\right.\nonumber\\
& \left.+\frac{1}{2\pi}\sum_{n=-\infty}^{\infty}\left\{ 2\pi\left|n+\theta\right|\text{Li}_{2}(e^{-2\pi\left|n+\theta\right|},2\pi\theta)+\text{Li}_{3}(e^{-2\pi\left|n+\theta\right|},2\pi\theta)\right\} \right].
\end{align}
Next, we can use that for periodic boundary conditions $\text{Li}_{s}(r,0)=\text{Li}_{s}(r)$,
and for antiperiodic $\text{Li}_{s}(r,\pi)=\text{Li}_{s}(-r)$. Finally,
using the definition of the polylogarithm $\text{\ensuremath{\text{Li}_{s}}}\left(z\right)=\sum_{p=1}^{\infty}\frac{z^{p}}{p^{s}}$
we can do the geometric sums in $n$ using
\beq
\sum_{n=-\infty}^{\infty}\left|n\right|(e^{-2\pi p})^{\left|n\right|}=\frac{2e^{-2\pi p}}{\left(1-e^{-2\pi p}\right)^{2}}=\frac{1}{2}\text{csch}^{2}\pi p
\eeq
\beq
\sum_{n=-\infty}^{\infty}(e^{-2\pi p})^{\left|n\right|}=\frac{1+e^{-2\pi p}}{1-e^{-2\pi p}}=\coth\pi p
\eeq
for the periodic case and
\beq
\sum_{n=1}\left|2n+1\right|(e^{-2\pi p})^{\left|2n+1\right|}=\frac{2\left(e^{-6\pi p}+e^{-2\pi p}\right)}{\left(1-e^{-4\pi p}\right)^{2}}=\text{csch}2\pi p\coth2\pi p=\frac{1}{4}\text{csch}^{2}\pi p+\frac{1}{4}\text{sech}^{2}\pi p
\eeq
\beq
\sum_{n=1}(e^{-2\pi p})^{\left|2n+1\right|}=\frac{2e^{-2\pi p}}{1-e^{-4\pi p}}=\text{csch}2\pi p
\eeq
for the antiperiodic case. For both cases we have been able to reduce the result to one single summation. The final step is to rewrite these sums in terms of known ones. For this purpose we rely on identities such as:
\beq
\frac{1}{2\pi}\sum_{p=1}^{\infty}\frac{\coth\pi p}{p^{3}}=\frac{7\pi^{2}}{360}
\eeq
which can be found in \cite{Ramanujan}. We have checked numerically the validity of the identities we use with arbitrary precision, but we could not find a reference in the literature where they were proven. The massless Casimir function for periodic boundary conditions gives:
\begin{align}
V_{\mathcal{C}}\left[a,0\right] & =\frac{1}{\left(2\pi a\right)^{2}}\left[\frac{1}{\pi^{2}}\sum_{p=1}^{\infty}\frac{1}{p^{4}}+\frac{1}{2\pi}\sum_{n=-\infty}^{\infty}\left\{ 2\pi\left|n\right|\text{Li}_{2}(e^{-2\pi\left|n\right|})+\text{Li}_{3}(e^{-2\pi\left|n\right|})\right\} \right] \nonumber\\
& =\frac{1}{\left(2\pi a\right)^{2}}\left[\frac{1}{\pi^{2}}  \text{Li}_{4}(1)+\frac{1}{2\pi}\sum_{n=-\infty}^{\infty}\sum_{p=1}^{\infty}\left\{ \frac{2\pi}{p^{2}}\left|n\right|(e^{-2\pi p})^{\left|n\right|}+\frac{1}{p^{3}}(e^{-2\pi p})^{\left|n\right|}\right\} \right]\nonumber\\
& =\frac{1}{\left(2\pi a\right)^{2}}\left[\frac{\pi^{2}}{90}+\frac{1}{2\pi}\sum_{p=1}^{\infty}\left\{ 2\pi\frac{1}{2p^{2}\sinh^{2}\pi p}+\frac{\coth\pi p}{p^{3}}\right\} \right]\nonumber\\
& =\frac{1}{\left(2\pi a\right)^{2}}\left[\frac{\pi^{2}}{90}+\frac{1}{\pi}\sum_{p=1}^{\infty}\left\{ 2\pi\frac{1}{p^{2}\left(\cosh\pi p-1\right)}\right\} +\frac{7\pi^{2}}{360}\right]\nonumber\\
& =\dfrac{1}{(2\pi a)^2}\dfrac{\mathcal{G}}{3},
\end{align}
where $\mathcal{G}\simeq 0.915966$ is Catalan's constant. For the case of antiperiodic boundary conditions the Casimir energy reads:
\begin{align}
V_{\mathcal{C}}\left[a,\frac{1}{2}\right] & =\frac{1}{\left(2\pi a\right)^{2}}\left[\frac{1}{\pi^{2}}\sum_{p=1}^{\infty}\frac{(-1)^{p}}{p^{4}}+\frac{1}{2\pi}\sum_{n=-\infty}^{\infty}\left\{ \pi\left|2n+1\right|\text{Li}_{2}(-e^{-\pi\left|2n+1\right|})+\text{Li}_{3}(-e^{-\pi\left|2n+1\right|})\right\} \right]\nonumber\\
& =\frac{1}{\left(2\pi a\right)^{2}}\left[\frac{1}{\pi^{2}} \text{Li}_{4}(-1)+\frac{1}{2\pi}\sum_{n=1}^{\infty}\left\{ \frac{\left(-1\right)^{p}\pi}{p^{2}}\left|2n+1\right|(e^{-\pi p})^{\left|2n+1\right|}+\frac{\left(-1\right)^{p}}{p^{3}}(e^{-\pi p})^{\left|2n+1\right|}\right\} \right] \nonumber\\
& =\frac{1}{\left(2\pi a\right)^{2}}\left[-\frac{7}{8}\frac{\pi^{2}}{90}-\frac{1}{2\pi}\sum_{p=1}^{\infty}\left\{ 2\pi\frac{\frac{(-1)^{p}}{4}\text{csch}^{2}\pi p+\frac{(-1)^{p}}{4}\text{sech}^{2}\pi p}{2p^{2}}+\frac{(-1)^{p}\text{csch}2\pi p}{p^{3}}\right\} \right]\nonumber\\
& =\dfrac{-1}{(2\pi a)^2}\dfrac{\mathcal{G}}{6}=-\dfrac{1}{2} V_{\mathcal{C}}\left[a,0\right].
\end{align}

In the ultraviolet region of supersymmetric models, all particles are essentially massless, so the sign of the total casimir potential is determined by the next order in the expansion. Retaining the next order in the polylogarithms we find  the following cuadratic terms in $am$. For the periodic case we have:
\begin{align}
V_{{\cal C}}^{(2)}\left[a,m,0\right]= & \left(am\right)^{2}\left\{ 2\pi\left(\log\left(2\pi ma\right)-\frac{1}{4}\right)-\text{Li}_{2}(1)+\sum_{n=1}^{\infty}2\pi\log\left(1-e^{-2\pi n}\right)\right\} \nonumber\\
= & \left(am\right)^{2}\left\{ \pi\log\left(2\pi ma\right)-\frac{\pi^{2}}{6}-\frac{5\pi}{2}+\log\frac{\Gamma\left(\frac{1}{4}\right)}{2\pi^{3/4}}\right\} \label{eq:cuadraticperiodic}
\end{align}
Finally, for the antiperiodic case we find:
\begin{align}
V_{{\cal C}}^{(2)}\left[a,m,\frac{1}{2}\right]= & \left(am\right)^{2}\left\{ -\text{Li}_{2}(-1)+\sum_{n=-\infty}^{\infty}\pi\log\left(1+e^{-\pi\left|2n+1\right|}\right)\right\}\nonumber\\
= & \left(am\right)^{2}\left\{ -\frac{\pi^{2}}{12}+\frac{3\pi}{4}\log2\right\}  \label{eq:cuadraticantiperiodic}
\end{align}
The sign of the second order terms is opposite to the leading order sign. This means that it is positive for bosons and negative for fermions. The sign of the potential is controlled by the most massive particles in the spectra. If the fermionic degrees of freedom win, the potential will be negative and it will develop a runaway solution.

\section{Consistency of the expansion about flat background in the non-compact dimensions}
\label{app:minkowski}

In this appendix, we show that taking the Minkowski metric as the background metric for the calculation of the one-loop potential of the 3D and 2D theories is a valid approximation up to the energy scales that we have studied in this paper. When computing the one-loop quantum effective action, one obtains, in the general case, extra contributions from the curvature of spacetime that take the form of quadratic terms in the Riemann tensor, the Ricci tensor and the Ricci scalar. In particular, since we want to study the vacua of the theory, we are interested in the cases in which the background metric is  dS, Minkowski or AdS. The typical energy scale associated to these terms is then given by the inverse of the radius of curvature, that is, the dS or AdS length. Hence, neglecting them (i.e. taking the background metric to be Minkowski) is justified as long as this  energy scale stays well below the energy scales of the other contributions to the potential. In our case, the typical energy scale associated with the effective potential is the KK scale of the compactification and in this way we can say that our calculations are consistent and meaningful as long as we fulfill the condition
\begin{equation}
\label{minksafe}
l_{\text{AdS/dS}}\gg l_{KK}.
\end{equation}
Note that it is by no means the intention of this appendix to give a detailed calculation of the contributions to the one-loop effective action coming from the non-zero curvature of the background metric but just to show that they are negligible for the cases we have studied. Before going to the two cases that are relevant for us, let us recall the general relation between the Ricci scalar and the radius of curvature (i.e. the dS or AdS length), $l$, which takes the form
\begin{equation}
\label{curvaturescalar}
\mathcal{R}_{(d)}=\pm \dfrac{d (d-1)}{l^2},
\end{equation}
where the positive sign corresponds to dS whereas the negative one corresponds to AdS.

{\it 3D potential}

In the circle compactification, we can use Einstein's equations to relate the Ricci scalar  with the 3D cosmoical constant, $\Lambda_{(3)}$, which at the same time is related with the potential, $V$, obtaining
\begin{equation}
\mathcal{R}_{(3)}=6 \Lambda_{(3)}=6\dfrac{V}{M_{p}^{(3)}},
\end{equation}
where $M_{p}^{(3)}=\sqrt{2\pi R}M_p$ is the 3D reduced Planck mass. In this way, since both the Planck mass and the potential depend on the radion, we can use Eq.  \eqref{curvaturescalar} to obtain an expression for $l$ as a function of the radion field and from there work out which values of $R$ fulfill the condition \eqref{minksafe}. This expression takes the form
\begin{equation}
\label{radius3d}
l=\dfrac{\sqrt{2\pi R}M_p^{(4)}}{V^{1/2}}.
\end{equation}
In particular, it is interesting to rewrite this equation in terms of quantities that we can easily obtain from our plots. For that purpose we recall that in our figures we plot $R^6 V$ in units of degrees of freedom, so  $n=720 \pi R^6 V$, the effective degrees of freedom, is the quantity that we read from the plots. In terms of these variables, Eq.  \eqref{radius3d} can be re-expressed as
\begin{equation}
l=\sqrt{\dfrac{720 \pi \times 2 \pi}{n r^3}}M_p^{(4)} R^{7/2},
\end{equation}
where, $r$ is just an arbitrary energy scale that ensures the right units for the potential, so we can take it to be $1$ GeV without loss of generality. From the figures we can safely say that, in these units, the potential takes values from $n=10$ to $n=70$  in all the cases we have studied (see Figs. \ref{majofig}-\ref{segment_periodic_SM}), so we can  take it to be $n\sim100$ to be safe. After plugging in the numbers, one obtains that taking Minkowski as the 3D background metric in the one-loop quantum effective action calculation is justified as long as
\begin{equation}
R^{-1}\ll 10^7 \ \mathrm{GeV},
\end{equation}
as in all our models. 

{\it 2D potential}

Due to conformal invariance of 2D gravity, we have seen that we cannot go to the Einstein frame in order to define a canonical potential or cosmological constant in 2D. However, the equations of  motion for the  $T^2/Z_4$ orbifold allow us to express the 2D Ricci scalar in terms of the potential as follows
\begin{equation}
\mathcal{R}_{(2)}=\dfrac{1}{(2 \pi M_p)^{2}}\dfrac{\partial_a V}{a}.
\end{equation}
As before, we can use Eq. \eqref{curvaturescalar} to obtain a relation between the curvature radius and the potential and it takes the form
\begin{equation}
l=2^{3/2} \pi M_p\left(\dfrac{a}{\partial_a V}\right).
\end{equation}
Following the same reasoning as in the 3D case, we want to express this in terms of variables that we can easily read from our plots, that is, in terms of $n=12 \pi^2 a^2 V/\mathcal{G}$ and its derivatives with respect to $\mathrm{\log}(a^{-1}\mathrm{[GeV]})$. Taking all this into account, we can rewrite the previous expression as
\begin{equation}
l^2=\dfrac{8 \pi^2 M_p^2}{\mathrm{\log} \ e} \ \dfrac{ \mathcal{G}}{12 \pi^2} \ \dfrac{a^4}{2 a^2 V + \partial_{\mathrm{\log}(a^{-1})}(a^2V)}.
\end{equation}
From  Figs. \ref{SMAllperiodic}-\ref{SUSYconstraints2} it can be seen that $n\sim 100$ is again a safe value  and that the slopes can also be overestimated  by taking them to be $\sim 100$. Plugging in these values, one obtains that 2D Minkowski background is a safe approximation if we fulfill 
\begin{equation}
a^{-1} \ll 10^{18} \ \mathrm{GeV}.
\end{equation}
This again implies that, with a conservative estimation, the potential we have considered would be changed so slightly by the corrections coming from a non-vanishing 2D curvature that we can safely ignore them.

\section{Example of MSSM spectrum}
\label{app:MSSMspectrum}

Here we just show a table with  the choice of SUSY masses taken to construct the figures related to the
SUSY case  in the text.  This particular choice is taken from \cite{Aparicio:2012iw}.
Note that the general results depend very weakly on the choice of spectrum.
\begin{table}
\begin{centering}
\begin{tabular}{|c|c|c|}
\hline 
Particle & Mass (TeV) & $(-1)^{(2s_{p}+1)}n_{p}$\tabularnewline
\hline 
\hline 
$\tilde{g}$ (gluinos) & 3.013 & +16\tabularnewline
\hline 
$\tilde{G}$ (gravitino) & 0.3-1 & 2\tabularnewline
\hline 
$\tilde{u_{L}}$,$\tilde{d}_{L}$,$\tilde{s}_{L}$,$\tilde{c}_{L}$ & 2.876 & -24\tabularnewline
\hline 
$\tilde{u_{R}}$,$\tilde{d}_{R}$,$\tilde{s}_{R}$,$\tilde{c}_{R}$ & 2.76 & -24\tabularnewline
\hline 
$\tilde{t}_{L}$ & 2. & -6\tabularnewline
\hline 
$\tilde{b}_{L}$ & 2.3 & -6\tabularnewline
\hline 
$\tilde{t}_{R}$,$\tilde{b}_{R}$ & 2.4 & -12\tabularnewline
\hline 
$\tilde{L}_{L}$ & 1.3 & -$2\times5$ \tabularnewline
\hline 
$\tilde{L}_{R}$ & 1.119 & -$2\times5$ \tabularnewline
\hline 
$\tilde{\tau}_{L}$ & 0.65 & -2\tabularnewline
\hline 
$\tilde{\tau}_{R}$ & 1.2 & -2\tabularnewline
\hline 
$\chi_{1}^{0}$ & 0.55 & 2\tabularnewline
\hline 
$\chi_{2}^{0}$ & 1.153 & 2\tabularnewline
\hline 
$\chi_{3}^{0}$,$\chi_{4}^{0}$ & 1.9 & 4\tabularnewline
\hline 
$\chi_{1}^{+}$ & 1.153 & 4\tabularnewline
\hline 
$\chi_{2}^{+}$ & 1.9 & 4\tabularnewline
\hline 
$A^{0}$ & 1.592 & -1\tabularnewline
\hline 
$\mathcal{H}$ & 1.6 & -1\tabularnewline
\hline 
$\mathcal{H}^{+}$ & 1.6 & -2\tabularnewline
\hline 
\end{tabular}\caption{Example of SUSY  spectrum taken from \cite{Aparicio:2012iw}.}
\par\end{centering}
\end{table}

\end{document}